\documentclass[prd,notitlepage,longbibliography,nofootinbib,superscriptaddress,onecolumn,preprintnumbers]{revtex4-2}
\usepackage[utf8]{inputenc}

\usepackage{bm}
\usepackage{comment} 
\usepackage[colorlinks=true,urlcolor=blue,anchorcolor=black,citecolor=blue,linkcolor=red,filecolor=black,menucolor=black,pagecolor=black,linktocpage=true,pdfproducer=medialab,pdfa=true]{hyperref}
\usepackage{graphicx}
\usepackage{amsmath,latexsym,amssymb,mathrsfs,ascmac,mathtools}
\usepackage{multirow}
\usepackage{braket}
\usepackage{placeins}
\usepackage{float}
\usepackage{subcaption}
\usepackage{makecell}
\allowdisplaybreaks
\begin{document}

\title{On the (In)equivalence of de Sitter and Holographic Entanglement Entropies in Lovelock Gravity}

\author{Takamasa Kanai}
\email{kanai@kochi-ct.ac.jp}

\affiliation{Department of Social Design Engineering,
National Institute of Technology (KOSEN), Kochi College,
200-1 Monobe Otsu, Nankoku, Kochi, 783-8508, Japan}

\begin{abstract}
We investigate the relation between de Sitter entropy and holographic entanglement entropy in higher-curvature gravity and its braneworld realization. Using the canonical formulation of the gravitational action and the Euclidean gravitational path integral, we derive the de Sitter entropy, including contributions from higher-curvature surface terms, and compare it with the holographic entanglement entropy obtained from the corresponding entropy functional.

We first consider Gauss-Bonnet gravity and show that the two entropies coincide for static asymptotically de Sitter braneworld spacetimes on the RS II model, while they generally differ for stationary spacetimes. The mismatch in the stationary case arises from an extrinsic-curvature contribution associated with the constant-time surface, which vanishes for static configurations but is generally nonzero for stationary geometries.

We then extend the analysis to general Lovelock gravity and show that the agreement between the de Sitter entropy and holographic entanglement entropy persists for static asymptotically de Sitter braneworld spacetimes in the RS II model beyond the Gauss-Bonnet case. We also comment that the same distinction between static and stationary configurations applies to braneworld black holes: the black hole entropy agrees with the holographic entanglement entropy in the static case, while the two generally differ for stationary braneworld black holes.

Our results clarify the relation between gravitational and holographic entropies in higher-curvature gravity and provide a broader perspective on holographic correspondence in Lovelock theories and their braneworld realizations.
\end{abstract}
\maketitle

\section{Introduction}

Entropy plays a fundamental role in gravitational physics and provides an important link between gravity, thermodynamics, and quantum theory. In particular, the entropy associated with horizons has been extensively studied since the discovery of the Bekenstein-Hawking area law. In Einstein gravity, the entropy of a black hole is proportional to the area of its event horizon \cite{Hawking:1974rv,Hawking:1975vcx,Bekenstein:1972tm,Bekenstein:1973ur,Bekenstein:1974ax}, while in higher-curvature theories of gravity the entropy generally receives additional contributions from curvature invariants and their associated boundary terms \cite{Wald:1993nt,Jacobson:1993vj,Jacobson:1993xs}. Understanding the origin and structure of gravitational entropy is therefore important, as gravitational entropy is expected to provide a key connection between gravity and quantum theory and to offer insights into the nature of quantum gravity.

A particularly important development in this context is the holographic description of quantum entanglement. The holographic principle, realized most explicitly through the AdS/CFT correspondence \cite{Maldacena:1997re,Gubser:1998bc,Witten:1998qj}, relates a gravitational theory in a higher-dimensional spacetime to a quantum field theory defined on its boundary. Within this framework, the entanglement entropy of a boundary subregion is encoded geometrically in the gravitational theory. For Einstein gravity, the Ryu--Takayanagi prescription identifies the holographic entanglement entropy with the area of a codimension-two extremal surface in the bulk \cite{Ryu:2006bv}. This relation was subsequently generalized to covariant settings by the Hubeny-Rangamani-Takayanagi prescription \cite{Hubeny:2007xt}. The holographic principle has also been investigated in braneworld settings, as studied in Ref.~\cite{Gubser:1999vj,Giddings:2000mu,Shiromizu:2001jm,Shiromizu:2001ve,Nojiri:2000eb,Nojiri:2000gb,Hawking:2000kj,Koyama:2001rf,Kanno:2002iaa}. Thus, holographic entanglement entropy provides a direct connection between quantum entanglement and the geometry of spacetime, making it a useful framework for investigating the relation between gravitational entropy and quantum information. The holographic principle has also been investigated in braneworld settings and applied to the study of the black hole information loss problem \cite{Almheiri:2019psf,Penington:2019npb}.

The holographic entropy functional is modified when the gravitational theory contains higher-curvature interactions. Among higher-curvature theories, Gauss--Bonnet gravity provides a particularly useful setting in which to investigate such questions. The Gauss-Bonnet term gives a nontrivial contribution to the gravitational dynamics in dimensions higher than four and modifies the entropy functional associated with gravitational horizons. In the context of holographic entanglement entropy, the corresponding functional is given by the Jacobson-Myers functional \cite{Jacobson:1993xs,Hung:2011xb}, which generalizes the area functional of Einstein gravity by including intrinsic curvature contributions of the codimension-two surface as well as an appropriate boundary term. This provides a natural framework for comparing gravitational horizon entropy with holographic entanglement entropy in higher-curvature gravity.

The relation between de Sitter entropy and holographic entanglement entropy was first investigated within Einstein gravity using de Sitter braneworlds (RS II model) \cite{Randall:1999vf} embedded in an anti-de Sitter (AdS) bulk. The de Sitter entropy in this construction is a horizon entropy obtained from the Euclidean gravitational action by evaluating the gravitational partition function in the semiclassical approximation \cite{Gibbons:1976ue}. In this setting, it was explicitly demonstrated that the de Sitter entropy of the braneworld agrees with the holographic entanglement entropy \cite{Iwashita:2006zj}. This correspondence was subsequently extended to higher-curvature gravity. In particular, it was shown that the same agreement holds for de Sitter braneworld solutions embedded in an AdS bulk in Gauss-Bonnet and Lovelock gravity \cite{Kushihara:2021fbr}. These results provide important evidence for a close relation between de Sitter entropy and holographic entanglement entropy beyond Einstein gravity. However, the analyses in these works were performed for specific braneworld configurations in which the de Sitter geometry is embedded in an AdS bulk. It is therefore natural to ask whether the equality between the two entropies is a more general property of gravitational spacetimes, independent of the particular braneworld construction.

In this work, we investigate the relation between de Sitter entropy and holographic entanglement entropy in more general gravitational spacetimes. In particular, we examine the relation between the two entropies for stationary geometries in higher-curvature gravity. We formulate the gravitational action in canonical form and analyze the relation between the entropies for stationary asymptotically de Sitter braneworld spacetimes in the RS II model. We find that the de Sitter entropy and holographic entanglement entropy generally differ for stationary geometries. On the other hand, when the stationary geometry is specialized to the static case, the two entropies coincide. We further extend this agreement in the static case to general Lovelock gravity, showing that the equality between the two entropies persists beyond Gauss-Bonnet gravity.

This distinction is important because stationary geometries can contain nontrivial shift-vector contributions and extrinsic-curvature terms that are absent in static configurations. Such terms can contribute to the boundary structure of the Euclidean gravitational action and therefore affect the relation between different entropy prescriptions. In particular, the extrinsic-curvature contribution vanishes for static configurations but generally remains nonzero for stationary geometries. Consequently, the equality between de Sitter entropy and holographic entanglement entropy need not persist once stationarity without staticity is allowed.

To investigate this issue, we formulate the Gauss-Bonnet action in canonical form and derive the de Sitter entropy from the Euclidean gravitational path integral, carefully retaining the relevant surface contributions. We then compare the resulting expression with the holographic entanglement entropy described by the Jacobson-Myers functional. We find that the two entropies generally differ for stationary spacetimes, while they coincide when the stationary geometry is specialized to the static case. This indicates that the agreement between the two entropies depends on the static nature of the geometry and does not generally extend to stationary configurations.

We also comment on the corresponding relation for braneworld black holes. In this case, the relevant gravitational entropy is the black hole entropy associated with the event horizon. We note that the same distinction between static and stationary configurations arises: the black hole entropy agrees with the holographic entanglement entropy in the static case, whereas the two generally differ for stationary black holes. This observation suggests that the static--stationary distinction is not specific to the de Sitter construction considered here, but reflects a more general feature of the relation between gravitational and holographic entropies in higher-curvature gravity.

Our results provide a broader perspective on the relation between gravitational and holographic entropies, and contribute to a deeper understanding of the conditions underlying their correspondence in higher-curvature gravity.

The remainder of this paper is organized as follows. In Sec.~\ref{sec:brane}, we introduce the Gauss-Bonnet braneworld setup and present the gravitational action relevant to our analysis. In Sec.~\ref{sec:cano}, we formulate Gauss-Bonnet gravity in the canonical formalism and discuss the associated boundary terms, which are required to derive the de Sitter entropy from the Euclidean gravitational action in the following section. In Sec.~\ref{sec:entropy}, we derive the de Sitter entropy from the Euclidean gravitational action and compare it with the holographic entanglement entropy obtained from the Jacobson-Myers functional. We show that the two entropy prescriptions generally differ for stationary spacetimes, while they coincide for static spacetimes. We then show that this difference arises from the contribution of the extrinsic curvature of constant-time hypersurfaces. In Sec.~\ref{sec:lovelock}, we extend our analysis to general Lovelock gravity and show that the two entropy prescriptions coincide for static spacetimes also in the general Lovelock case. Finally, in Sec.~\ref{sec:conclusion}, we discuss the physical implications of our results and conclude with some prospects for future investigations. In particular, we discuss the extension to braneworld black holes and show that, in this case as well, the two entropy prescriptions generally differ for stationary spacetimes but coincide for static spacetimes.

For clarity, the geometrical quantities used throughout this paper are summarized in Table~\ref{tab:geometry}.
\begin{table}[htbp]
\centering
\caption{A summart of notation}
\label{tab:geometry}
\begin{tabular}{c|c|c|c|c|c}
\hline
  & metric $\&$ index  & curvature & extrinsic curvature & unit normal vector & surface type \\
\hline
$M^+\cup M^-$ & $g_{MN}$ & $R_{MNKL}[g]$ & Nothing & Nothing & bulk\\
\hline
$\Sigma_{n}$ & $h_{ij}$ & $\tilde{R}_{ijkl}[h]$ & $H_{ij}[u]$ &$u_M$ & Constant-time slice \\
\hline
$M_{n-1}$ & $q_{\mu\nu}$ &${}^{(n-1)}R_{\mu\nu\rho\sigma}[q]$ & $K_{\mu\nu}[n]$ & $n_{M}$ & Brane \\
\hline
$\Sigma_{n-1}$ & $\gamma_{XY}$ & ${}^{(n-2)}R_{XYZW}[\gamma]$ & ${}^{(n-2)}K_{XY}[n]$ &$u_{M},n_{M}$ & Brane on a constant-time slice \\
\hline
$\Gamma^+\cup\Gamma^-$ & $G_{xy}$ & ${}^{(n-2)}\tilde{R}_{abcd}[G]$ & Not used & $u_{M},r_{M}$ & Cosmological horizon on a constant-time slice \\
\hline
$\partial\Gamma$ & $p_{AB}$ & ${}^{(n-3)}R_{ABCD}[p]$ & ${}^{(n-3)}K_{AB}[n]$ & $u_{M},n_{M},r_{M}$ & Brane–horizon intersection on a constant-time slice \\
\hline
\end{tabular}
\end{table}

For the extrinsic curvature, the quantity in square brackets specifies the normal vector with respect to which the extrinsic curvature is defined. Quantities not listed in the table are not used in this paper.

\begin{figure}[t]
    \centering
    \includegraphics[width=0.5\textwidth]{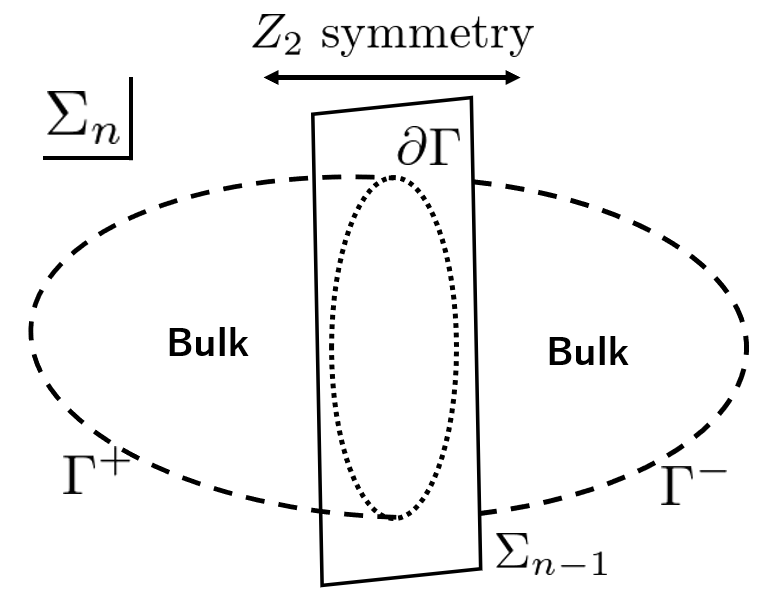}
    \caption{Schematic illustration of the RS II braneworld with a de Sitter brane at constant time.}
    \label{fig:spacetime}
\end{figure}

Figure~\ref{fig:spacetime} illustrates the spatial geometry on a constant-time slice. The central plane represents the $(n-1)$-dimensional brane, while the dashed curves represent the cosmological horizons $\Gamma^+ \cup \Gamma^-$ in the $n$-dimensional bulk spacetime. The wavy curve represents the cosmological horizon on the brane, denoted by $\partial\Gamma$.

We consider an RS~II braneworld in which an $(n-1)$-dimensional brane is embedded in an $n$-dimensional bulk spacetime. The geometry induced on the brane is assumed to be asymptotically de Sitter and therefore possesses a cosmological horizon $\partial\Gamma$. On a constant-time slice, this horizon is the intersection of the brane with the bulk horizon $\Gamma^+ \cup \Gamma^-$. The bulk horizon extends away from the brane and forms a hypersurface surrounding the brane, as illustrated in Fig.~\ref{fig:spacetime}. This configuration provides the geometric setting for the entropy calculation considered in this work.

In this paper, we use square brackets to indicate complete antisymmetrization over the enclosed indices. For a rank-$n$ tensor $A_{a_1\cdots a_n}$, we define
\begin{align}
A_{[a_1a_2\cdots a_n]}:=\frac{1}{n!}\sum_{\sigma\in S_n}\operatorname{sgn}(\sigma)A_{a_{\sigma(1)}a_{\sigma(2)}\cdots a_{\sigma(n)}},
\end{align}
where $S_n$ is the set of all permutations of $n$ indices, and $\operatorname{sgn}(\sigma)$ is the sign of the permutation $\sigma$, taking the value $+1$ for even permutations and $-1$ for odd permutations.

\section{Braneworld in the Gauss-Bonnet gravity}
\label{sec:brane}

Our purpose is to investigate whether the formula given in Ref.~\cite{Hung:2011xb} coincides with the de Sitter entropy in braneworlds (RS II model) with Gauss-Bonnet gravity. We consider $Z_2$-symemetric braneworld model in the anti-deSitter bulk. The system is composed of the $n$-dimesional bulk $(M_n^{\pm}, g_{MN})$ and the brane $(M_{n-1},q_{\mu\nu})$. The action of the Gauss-Bonnet gravity with boundary is given by \cite{Myers:1987yn,Maeda:2003vq,Cano:2018ckq}
\begin{align}
\label{GB action}
I=\frac{1}{16\pi G_n}\int_{M_n^{+}\cup M_n^-}d^nx\sqrt{-g}\left(R-2\Lambda+\frac{\beta l^2}{4}\mathcal{L}_{GB}\right)+\int_{M_{n-1}}d^{n-}x\sqrt{-q}\left(-\sigma+\frac{1}{16\pi G_n}[Q]^-\right),
\end{align}
where $G_n$ is the $n$-dimensional Newton constant, $R$ is the $n$-dimensional Ricci scalar, $\Lambda$ is a negative cosmological constant, $l$ is supposed to be the anti-deSitter curvature length and $\beta$ is a dimensionless constant. Here, the Gauss-Bonnet Lagrangian $\mathcal{L}_{GB}$ is given as
\begin{align}
\mathcal{L}_{GB}=R^2-4R_{MN}R^{MN}+R_{MNKL}R^{MNKL},
\end{align}
and the gravitational surface term $Q$ is written in
\begin{align}
Q=2K+\beta l^2(J-2\ ^{(n-1)}G_{\mu\nu}K^{\mu\nu}),
\end{align}
where ${}^{(n-1)}G_{\mu\nu}$ is the $(n-1)$-dimensional Einstein tensor, $J$ is the trace of $J_{\mu\nu}$ defiend by
\begin{align}
J_{\mu\nu}&=-\frac{1}{3}(2K_{\mu\rho}K_{\nu\sigma}K^{\rho\sigma}-2KK_{\mu\rho}K^\rho_{\ \ \nu}-K_{\mu\nu}K_{\rho\sigma}K^{\rho\sigma}+K^2K_{\mu\nu}),\\
J&=-\frac{1}{3}(2K^{\mu}_{\ \nu}K^{\nu}_{\ \rho}K^{\rho}_{\ \mu}-3KK_{\mu\nu}K^{\mu\nu}+K^3)\notag\\
&=-\frac{1}{3}q^{\mu_1\mu_1\mu_2}_{\ \rho_1\sigma_1\rho_2}K^{\rho_1}_{\ \ \mu_1}K^{\sigma_1}_{\ \ \nu_1}K^{\rho_2}_{\ \ \mu_2},\\
q^{\mu_1\nu_1\mu_2}_{\ \rho_1\sigma_1\rho_2}&:=3!q^{\mu_1}_{\ \ [\rho_1}q^{\nu_1}_{\ \ \sigma_1}q^{\rho_2}_{\ \ \mu_2]},
\end{align}
and $K_{\mu\nu}$ is the extrinsic curvature of $M_{n-1}$ (whose the normal direction is taken to outward for $M_{n}^+$). Supposing that the location of the brane $M_{n-1}$ is $y=0$ in the Gaussian normal coordinate $(y,x^{\mu})$ around the brane, $[F]^-$ is defined by
\begin{align}
[F]^-:=\lim_{y\rightarrow+0}F-\lim_{y\rightarrow-0}F.
\end{align}
Then, the bulk field equation is 
\begin{align}
G_{MN}+\Lambda g_{MN}+\frac{\beta l^2}{2}H_{MN}=0,
\end{align}
where 
\begin{align}
H_{MN}=RR_{MN}-2R_{MK}R^K_{\ \ N}-2R^{KL}R_{MKNL}+R_{MKLP}R_{N}^{\ \ KLP}-\frac{1}{4}g_{MN}\mathcal{L}_{GB}.
\end{align}
The junction condition is 
\begin{align}
\label{junction condition}
[K^{\mu}_{\ \nu}-\delta^{\mu}_{\ \nu}K]^-+\frac{\beta l^2}{2}[3J^{\mu}_{\ \nu}-\delta^\mu_{\ \nu}J-2P^\mu_{\ \rho\nu\sigma}K^{\rho\sigma}]^-=8\pi G_n\tau^{\mu}_{\ \nu},
\end{align}
where
\begin{align}
P_{\mu\rho\nu\sigma}&=^{(n-1)}R_{\mu\rho\nu\sigma}-2\ ^{(n-1)}R_{\mu[\nu}q_{\sigma]\rho}+2\ ^{(n-1)}R_{\rho[\nu}q_{\sigma]\mu}+\ ^{(n-1)}Rq_{\mu[\nu}q_{\sigma]\rho},
\end{align}
and $\tau_{\mu\nu}$ is the energy momentum tensor on the brane. Since we focus on the vacuum brane, $\tau_{\mu\nu}=-\sigma q_{\mu\nu}$, where $\sigma$ is the brane tension.

In the following, we first formulate Gauss-Bonnet gravity in the canonical formalism. In Sec.~\ref{sec:entropy}, we then derive the de Sitter entropy from the canonical formulation for general stationary asymptotically de Sitter braneworld spacetimes in the RS II model and compare it with the holographic entanglement entropy. We show that the two entropies do not coincide for general stationary spacetimes because the extrinsic curvature of constant-time hypersurfaces does not vanish in general. In contrast, for static spacetimes, the extrinsic curvature of constant-time hypersurfaces vanishes, and consequently the de Sitter entropy coincides with the holographic entanglement entropy.

\section{Canonical form in Gauss-Bonet gravity}
\label{sec:cano}

The deSitter entropy is obtained by evaluating the action functional in a saddle point approximation of the Euclidean path integral. In this section, we introduce the canonical form in the Gauss-Bonnet gravity \cite{Padilla:2003qi} to compute the deSitter entropy. The Gauss-Bonnet action (\ref{GB action}) is most elegantly written in terms of differential forms. The metric of $n$-dimensional spacetime $M$ is given by
\begin{align}
g_{ab}=\eta_{ij}(E^i)_a\otimes (E^j)_b,
\end{align}
where $\lbrace E^i\rbrace$ is an orthogonal basis of $1$-forms, and indices are raised/lowered using $\eta_{ij}={\rm diag}(-1,1,\cdots,1)$. We write $\lbrace X_i\rbrace$ for the dual basis of vectors.

We will find it useful to define the following forms,
\begin{align}
e_{i_1\cdots i_m}=\frac{1}{(d-m)!}\epsilon_{i_1\cdots i_mi_{m+1}\cdots i_d}E^{i_{m+1}}\land\cdots\land E^{i_d},
\end{align}
where $\epsilon_{i_1\cdots i_mi_{m+1}\cdots i_d}$ is the totally antisymmetric tensor with $\epsilon_{0\cdots d-1}=1$. Notice that the scalar $d$-form, $e$ is the volume element on $M$.

The curvature $2$-form is given by 
\begin{align}
\Omega^{i}_{\ j}=d\omega^i_{\ j}+\omega^i_{\ k}\land\omega^k_{\ j}=R^i_{\ jkl}E^k\land E^l,
\end{align}
where the right hand equation above gives the Riemann tensor, $R^A_{\ \ BCD}$ and $\omega^i_{\ j}$ is the connection $1$-form defiend by
\begin{align}
\omega^i_{\ j}&=\omega^i_{\ jk}E^k,\\
\omega^i_{\ jk}&=(E^i)^a(X_j)^b\nabla_a(X_k)_b.
\end{align}
The Gauss-Bonnet action (\ref{GB action}) can now be written by
\begin{align}
I&=\frac{1}{16\pi G}\int_M\Omega^{ij}\land e_{ij}-2\Lambda e+\frac{\beta l^2}{4}\Omega^{ij}\land \Omega^{kl}\land e_{ijkl}\notag\\
&\ \ \ -\frac{1}{16\pi G}\int_{\partial M}\theta^{ij}\land e_{ij}+\frac{\beta l^2}{2}\theta^{ij}\land \left(\Omega^{kl}-\frac{2}{3}\theta^k_{\ m}\land\theta^{ml}\right)\land e_{ijkl},
\end{align}
where $\theta^{ij}$ is the second fundamental form defined by
\begin{align}
\theta^{ij}=\omega^{ij}-\omega_0^{ij}.
\end{align}
Here, $\omega^{ij}$ is the connection of the produc metric $g_0$ which agrees with the metric on the boundary and, we have made use of the following identity
\begin{align}
\label{identity1}
E^i\land e_{i_1\cdots i_m}&=\delta^i_{\ i_m}e_{i_1\cdots i_{m-1}}-\delta^i_{\ i_{m-1}}e_{i_1\cdots i_{m-2}i_m}+\cdots+(-1)^{m-1}\delta^i_{\ i_1}e_{i_1\cdots i_m}.
\end{align}
First, we choose a timelike vector field $\partial/\partial t$. We introduce a spacelike hypersurface $\lbrace\Sigma\rbrace$ which is the $t=$const surface and assume that the hypersurface meets the timelike part of the boundary $\partial M=B$ orthogonally.

We can write the metric of $M$ in ADM form
\begin{align}
g=-N^2dt^2+h_{ab}(dx^a+N^adt)(dx^b+N^bdt),
\end{align}
where $N$ is the lapse function, $N^a$ is the shift vector, $h_{ab}$ is the induced metric on$\Sigma$, and $\lbrace a,b,c\dots\rbrace$ represents indices of $\Sigma$. The orthogonal basis $\lbrace E^i\rbrace=E^\perp\cup\lbrace E^a\rbrace$ of $1$-forms follow as
\begin{align}
E^\perp=Ndt,\ \ \ E^a=E^a_{\ b}(dx^b+N^bdt).
\end{align}
The dual basis of vectors is given by 
\begin{align}
X_\perp=\frac{1}{N}\left(\frac{\partial}{\partial t}-N^a\frac{\partial}{\partial x^a}\right),\ \ \ X_a=h_a^{\ b}\frac{\partial}{\partial x^a}.
\end{align}

In this paper, we will employ the orthogonal frames and the differential forms, and describe the decompoition of the bulk Riemann tensor into spatial tensors on $\Sigma$ using the Gauss-Codazzi equations. At first, we introduce the decomposition of the connection $\omega^{ab}$.
\begin{align}
\omega^{\perp a}&=H^a+a^aE^\perp,\\
\omega^{ab}&=\tilde{\omega}^{ab}+l^{ab}E^\perp.
\end{align}
Here, the $1$-form $H^a=H^a_{\ b}E^b$, where $H_{ab}$ is the extrinsic curvature of $\Sigma$ and $\tilde{\omega}^{ab}$ is the connection for the induced metric $h_{ab}$. The $0$-form $a^a$ is a vector and the $0$-form $l^{ab}$ is an sntisymmetric tensor. This gives the following condition
\begin{align}
a_aE^a&=\frac{\tilde{d}N}{N},
\label{a}
\end{align}
where $\tilde{d}$ is the exterior derivative operator.

In differential form, the Gauss-Codazzi equations are written as follows 
\begin{align}
\Omega^{\perp a}&=\tilde{\nabla}H^a+E^\perp\land\Bigl[\$_\perp H^a-\frac{1}{N}\tilde{\nabla}(Na^a)+l^{ab}H_b\Bigr],\\
\Omega^{ab}&=\tilde{\Omega}^{ab}+H^a\land H^b+E^\perp\land\Bigl[\$_\perp\tilde{\omega}^{ab}-\frac{1}{N}\tilde{\nabla}(Nl^{ab})-H^aa^b+H^ba^a\Bigr],
\end{align}
where $\tilde{\Omega}^{ab}=d\tilde{\omega}^{ab}+\tilde{\omega}^a_{\ c}\tilde{\omega}^{cb}$ is the curvature $2$-form for $\Sigma$, $\tilde{\nabla}$ is the covariant exterior derivative operator on $\Sigma$, and $\$_\perp$ is the Lie derivative with respect to the vector $X_\perp$ on $\Sigma$.

We find that the bulk action is given by
\begin{align}
I_{{\rm bulk}}=I_{{\rm kinetic}}+I_1+I_2+I_3,
\end{align}
where
\begin{align}
I_{{\rm kinetic}}&=\frac{1}{16\pi G_n}\int_ME^\perp\land\Bigl\{2\$_\perp H^a\land\xi_a\notag\\
&\qquad\qquad\qquad\qquad+\beta l^2\bigl[\tilde{\nabla}H^a\land\$_\perp\tilde{\omega}^{bc}+\$_\perp H^a\land F^{bc}\bigr]\land\xi_{abc}\Bigr\},\\
I_1&=\frac{1}{16\pi G_n}\int_M E^\perp\land\Bigl\{2\Lambda\xi-F^{ab}\land\xi_{ab}-\frac{\beta l^2}{4}F^{ab}\land F^{cd}\land\xi_{abcd}\Bigr\},\\
I_2&=\frac{1}{16\pi G_n}\int_M E^\perp\land\Bigl\{-\frac{2}{N}\tilde{\nabla}(Na^a)\land\xi_a\notag\\
&\qquad\qquad\qquad\qquad-\beta l^2\Bigl[\frac{1}{N}\tilde{\nabla}(Na^a)\land F^{bc}+2\tilde{\nabla}H^a\land H^{[b}a^{c]}\Bigr]\land\xi_{abc}\Bigr\},\\
I_3&=-\frac{1}{16\pi G_n}\int_M E^\perp\land\Bigl\{\beta l^2\Bigl[\tilde{\nabla}H^a\land\frac{1}{N}\tilde{\nabla}(Nl^{bc})-l^{ad}H_d\land F^{bc}\Bigr]\land\xi_{abc}\Bigr\},
\end{align}
with
\begin{align}
F^{ab}=\tilde{\Omega}^{ab}+H^a\land H^b.
\end{align}
In deriving the above equations, we used the following relation:
\begin{align}
l^{ab}H_b\land\xi_a=0.
\end{align}

Note that we have introduced the analogue of $e_{A_1\cdots A_m}$ on $\Sigma$,
\begin{align}
\xi_{a_1\cdots a_m}=e_{\perp a_1\cdots a_m}=\frac{1}{(n-1-m)!}\epsilon_{a_1\cdots a_ma_{m+1}\cdots a_{n-1}}E^{a_{m+1}}\land\cdots\land E^{a_{n-1}}.
\end{align}
The kinetic term is given by
\begin{align}
\label{kinetic}
I_{{\rm kinetic}}&=\frac{1}{16\pi G_n}\int_ME^\perp\land\Bigl\{-2\$_\perp H^a\land\xi_a\notag\\
&\qquad\qquad\qquad\qquad+\beta l^2\bigl[\tilde{\nabla}H^a\land\$_\perp\tilde{\omega}^{bc}+\$_\perp H^a\land F^{bc}\bigr]\land\xi_{abc}\Bigr\}\notag\\
&=\frac{1}{16\pi G_n}\int_ME^\perp\land\Bigl\{\pi^a\land\$_\perp E^a\notag\\
&\qquad\qquad\qquad\qquad-\beta l^s\bigl[H^a\land\$_\perp\tilde{\Omega}^{bc}-\tilde{\nabla}H^a\land\$_\perp\tilde{\omega}^{bc}\bigr]\land\xi_{abc}\Bigr\},\notag\\
\end{align}
where 
\begin{align}
\pi_a&=2\xi_{ab}\land H^b+\beta l^2\xi_{abcd}\land H^b\land\Lambda^{cd},\\
\Lambda^{ab}&=\tilde{\Omega}^{ab}+\frac{1}{3}H^a\land H^b,
\end{align}
and the following was used to transform the equation,
\begin{align}
\$_\perp H^a\land H^b\land H^c\land\xi_{abc}=\frac{1}{3}\$_\perp(H^a\land H^b\land H^c)\land\xi_{abc},
\end{align}
Moreover, $I_{{\rm kinetic}}$ becomes
\begin{align}
&\int_ME^\perp\land\Bigl[H^a\land\$_\perp\tilde{\Omega}^{bc}-\tilde{\nabla}H^a\land\$_\perp\tilde{\omega}^{bc}\Bigr]\land\xi_{abc}\notag\\
&=\int dt\int_\Sigma\Bigl[H^a\land\tilde{\nabla}(N\$_\perp\tilde{\Omega}^{bc})-\tilde{\nabla}H^a\land N\$_\perp\tilde{\omega}^{bc}\Bigr]\land\xi_{abc}\notag\\
&=-\int dt\int_\Sigma\tilde{d}\Bigl[NH^a\land\$_\perp\tilde{\omega}^{bc}\land\xi_{abc}\Bigr]\notag\\
&=\int dt\int_{S_t}NH^a\land\$_\perp\tilde{\omega}^{bc}\land\xi_{abc},
\end{align}
where $S_t$ is the spacelike surface on $\Sigma$, and in the first equality, we used the relation
\begin{align}
E^\perp\land\$_\perp\tilde{\Omega}^{ab}=dt\land\tilde{\nabla}(N\$_\perp\tilde{\omega}^{ab}),
\end{align}
in the second equality, we used the fact that $\tilde{\nabla}\xi_{abc}=0$ and, in the last equality, we used the Stokes theorem. Therefore, Eq. (\ref{kinetic}) is 
\begin{align}
I_{{\rm kinetic}}=\frac{1}{16\pi G_n}\int dt\int_\Sigma\pi_a\land N\$_\perp E^a+\frac{1}{16\pi G_n}\int dt\int_{S_t}\beta l^2NH^a\land\$_\perp\tilde{\omega}^{bc}\land\xi_{abc}.
\end{align}
Here, $\dot{}$ is the derivative with resepect to $t$, the first term can stii be calculated from the definition of the Lie derivative,
\begin{align}
\int dt\int_\Sigma \pi_a\land\$_\perp E^a=\int dt\int_\Sigma N^a\mathcal{H}_a-\int dt\int_{S_t}(-1)^nN^a\pi_aE^b_{\ a},
\end{align}
where $\mathcal{H}_a$ is the momentum constraint, which is given by
\begin{align}
\label{pi}
\mathcal{H}_a=(-1)^{n-1}E^b_{\ a}\tilde{\nabla}\pi_b.
\end{align}
 In deriving Eq. (\ref{pi}), we used the facts that
\begin{align}
\$_\perp E^a&=\frac{1}{N}\left(\dot{E}^a-\$_{\vec{N}}E^a\right),\\
\$_{\vec{N}} E^a&=i_{\vec{N}}(\tilde{d}E^a)+\tilde{d}(i_{\vec{N}}E^a)=-N^ci_{\frac{\partial}{\partial x^c}}(\tilde{\omega}^a_{\ c})E^c+\tilde{\nabla}(E^a_{\ b}N^b),\\
i_{\frac{\partial}{\partial x^c}}(\tilde{\omega}^a_{\ b})\pi_a\land E^b&=0.
\end{align}
Here, $i_YA$ is the interior product of the vector $Y$ and the $p$-form $A$ and, the third equation is derived from $i_{\frac{\partial}{\partial x^c}}\tilde{\omega}^{ab}$ is antisymetric tensor and the fact that
\begin{align}
\pi^a\land E^b=\pi^b\land E^a.
\end{align}
This is not obvious but can be found using the symmetriies of $H^{ab}$ and the Riemann tensor.

The remaining terms in the bulk action can be written as follows:
\begin{align}
I=I_{{\rm bulk}}+I_{{\rm bdy}},
\end{align}
where
\begin{align}
\label{bulk}
I_{{\rm bulk}}&=\frac{1}{16\pi G_n}\int dt\int_\Sigma\pi_a\land\dot{E}^a-N\mathcal{H}-N^a\mathcal{H}_a,
\end{align}
\begin{align}
\label{bdy}
I_{{\rm bdy}}&=\frac{1}{16\pi G_n}\int dt\int_{S_t}(-1)^nN^a\pi_bh^b_{\ a}+N\Bigl\{2a^b\xi_b+\beta l^2\Bigl[-H^b\land\$_\perp\tilde{\omega}^{cd}+a^bF^{cd}+\tilde{\nabla}H^bl^{cd}\Bigr]\land\xi_{bcd}\Bigr\}\notag\\
&\qquad+S_{{\rm GH}},\\
\mathcal{H}&=2\Lambda\xi-F^{ab}\land\xi_{ab}-\frac{\beta l^2}{4}F^{ab}\land F^{cd}\land\xi_{abcd}.
\end{align}
Note that $\mathcal{H}$ is the Hamiltonian constraint and, in the deriving the expression (\ref{bulk}) and Ea. (\ref{bdy}) for the boundary term, we make use of the Bianchi identity and the following:
\begin{align}
H_a\land\left[l^{ba}F^{cd}+l^{cd}\tilde{\Omega}^{ba}\right]\land\xi_{bcd}=0.
\end{align}
We find from Eq. (\ref{identity1}) that this identity holds true
\subsection{Action on the boundary}
In this subsection, we derive the gravitational action on the boundary. We need to understand how bulk quantities project on $S_t$. In this paper, we assume that the foliation of $B_t$ can be give throughout the spacetime $M$. At first, the metric can be written in the Gaussian normal coordinates
\begin{align}
g=dz^2-E^\perp\otimes E_\perp+\delta_{AB}E^A\otimes E^B,
\end{align}
where the indice $A,B,C,cdots$ represents the label components in $S_t$. Note that near $B$, $E^z_{\ a}=\delta^z_{\ a}$ and $N^z=0$.

Also, we find that the components of the connnection is given by 
\begin{align}
\omega^{\perp z}&=b_AE^A+cE^\perp,\\
\omega^{\perp A}&=\hat{H}^A+a^AE^\perp+b^AE^z,\\
\omega^{zA}&=-\hat{K}^A+b^AE^\perp,\\
\omega^{AB}&=\hat{\omega}^{AB}+l^{AB}E^\perp+\chi^{AB}E^z,
\end{align}
where $\chi^{AB}$ is an antisymmetric tensor and 
\begin{align}
c&=a^z,\\
b^A&=l^{zA}=H^{zA}.
\end{align}
Note that $\hat{}$ represents intrinsic quantities on $S_t$. Therefore, $\hat{\omega}^{AB}$ is the connection for the induced metric $\gamma_{AB}$ on $S_t$. We have also defined,
\begin{align}
\hat{H}^A&=\hat{H}^A_{\ B}E^B,\\
\hat{K}^A&=\hat{K}^A_{\ B}E^B,
\end{align}
where $\hat{H}^A_{\ B}$ and $\hat{K}^A_{\ B}$ are the extrinsic curvatures of $S_t$. We now use the fact that 
\begin{align}
\$_\perp\tilde{\omega}^{zA}=\hat{\$}_\perp\hat{K}^A+E^z\{\dots\},
\end{align}
to rewrite the remaing terms other than the Gibbons-Hawking term in Eq. (\ref{bdy}). We find that
\begin{align}
I_{{\rm bdy}}=&\frac{1}{16\pi G_n}\int dt\int_{S_t}(-1)^{n-1}N^a\pi_bh^b_{\ a}+2Nc\phi\notag\\
&+\beta l^2\Bigl[b_CE^C\land\$_\perp\tilde{\omega}^{AB}+cF^{AB}+F^zl^{AB}-2\hat{H}^A\land\hat{\$}_\perp\hat{K}^B\notag\\
&-2a^AF^{zB}-2F^Ab^B\Bigr]\land\phi_{AB}+S_{{\rm GH}},
\end{align}
where $F^a=\tilde{\nabla}H^a$. We have also introduced the $S_t$ analogue of $\xi_{a_1\cdots a_m}$ and $e_{i_1\cdots i_m}$:
\begin{align}
\phi_{i_1\cdots i_m}&=\xi_{zi_1\cdots i_m}=e_{\perp zi_1\cdots i_m}.
\end{align}
Meanwhile, terms like $F^z$ and $F^{AB}$ are given by
\begin{align}
F^z&=\hat{d}(b_AE^A)+\hat{H}^A\land\hat{K}_A,\\
F^A&=\hat{\nabla}\hat{H}-b_BE^B\land\hat{K}^A,\\
F^{zA}&=-\hat{\nabla}\hat{K}^{A}+b_BE^B\land\hat{H}^A,\\
F^{AB}&=\hat{\Omega}^{AB}+\hat{H}^A\land\hat{H}^B-\hat{K}^A\land\hat{K}^B,
\end{align}
where $\hat{\Omega}^{AB}$ is the intrinsic curvature two-form of the codimension-two spatial boundary $\partial S_t$. Note that the non-zero components of $\theta^{ij}$ are only $\theta^{zi}$ and $\theta^{\perp z}$. Then, the Gibbons-Hawking term can be written as
\begin{align}
I_{GH}=&\frac{1}{16\pi G_n}\int dt\int_{S_t}-2N\hat{K}^A\land\phi_A-2Nc\phi-\beta l^2N\hat{K}^A\land\Bigl[F^{BC}+\frac{2}{3}\hat{K}^B\land\hat{K}^C\Bigr]\land\phi_{ABC}\notag\\
&+\beta l^2\Bigl\{-2b^A\Bigl[F^B+\frac{2}{3}b_CE^C\land\hat{K}^B\Bigr]-2\hat{K}^B\land\Bigl[G^B+\frac{2}{3}\Bigl(c\hat{K}^B+b^Bb_CE^C\Bigr)\Bigr]\notag\\
&\qquad\ \ \ \ -c\Bigl[F^{AB}+\frac{2}{3}\hat{K}^A\land\hat{K}^B\Bigr]+b_CE^C\land\Bigl[G^{AB}-\frac{4}{3}b^{[A}\hat{K}^{B]}\Bigl]\Bigl\}\land\phi_{AB},
\end{align}
where 
\begin{align}
G^A&=\hat{\$}_\perp\hat{H}^A-\frac{1}{N}\hat{\nabla}(Na^A)+l^{AB}\hat{H}_B-c\hat{K}^A-b^Ab_BE^B,\\
G^{AB}&=\hat{\$}_\perp\hat{\omega}^{AB}-\frac{1}{N}\hat{\nabla}(Nl^{AB})-2\hat{H}^{[A}a^{B]}-2\hat{K}^{[A}b^{B]}.
\end{align}
Note that we drropped all terms like $E^z\{\dots\}$ as we are now integrating over $B$.
\begin{align}
I_{{\rm bdy}}=&\frac{1}{16\pi G_n}\int dt\int_{S_t}(-1)^{n-1}N^A\pi_BE^B_{\ A}-\Bigl\{2\hat{K}^A\land\phi_A\notag\\
&\qquad\qquad\qquad\qquad+\beta l^2\hat{K}^A\land\Bigl[\hat{\Omega}^{BC}-\hat{H}^B\land\hat{H}^C-\frac{1}{3}\hat{K}^B\land\hat{K}^C\Bigr]\land\phi_{ABC}\Bigr\}+I_{{\rm total}},
\end{align}
where
\begin{align}
\label{total derivative}
I_{{\rm total}}&=\frac{1}{16\pi G_n}\int dt\int_{S_t}\beta l^2N\Bigl\{\hat{d}(b_CE^C)l^{AB}-\frac{1}{N}b_CE^C\land\hat{\nabla}(Nl^{AB})\notag\\
&\qquad\qquad\qquad\qquad+2a^A\hat{\nabla}\hat{K}^B-\frac{2}{N}\hat{K}^B\land\hat{\nabla}(Na^A)\Bigr\}\land\phi_{AB}\notag\\
&=-\frac{1}{16\pi G_n}\int dt\int_{\partial S_t}\beta l^2N\Bigl\{b_CE^Cl^{AB}+2a^A\hat{K}^B\Bigr\}\land\phi_{AB}.
\end{align}
Here, we used the condition
\begin{align}
\hat{\nabla}\phi_{A\cdots N}=0,
\end{align}
and the identity
\begin{align}
\hat{H}^C\land\hat{K}_Cl^{AB}\land\phi_{AB}-2\hat{K}^A\land{H}_Cl^{BC}\land\phi_{AB}&=0,\\
2\hat{K}^A\land\hat{H}^Bl^C_{\ D}\land E^D\land\phi_{ABC}&=0.
\end{align}
We can find that these identities hold true from Eq. (\ref{identity1}). Note that the first term of the total derivative term on $S_t$ Eq. (\ref{total derivative}) vanishes on the cosmological horizon of the stationary spacetime. It is because of $b_C=H^z_{\ C}$.

Finally, the action including the boundary terms in the present setup is given by
\begin{align}
I=&-\frac{1}{16\pi G_n}\int dt\int_{\Sigma_t}d^{n-1}x(N\mathcal{H}+N^i\mathcal{H}_i)\notag\\
&+\frac{1}{16\pi G_n}\int dt\int_{\Gamma_t}(-1)^nN^a\pi_bh^b_{\ a}+N\Bigl\{2a^b\xi_b+\beta l^2\Bigl[-H^b\land\$_\perp\tilde{\omega}^{cd}+a^bF^{cd}+\tilde{\nabla}H^bl^{cd}\Bigr]\land\xi_{bcd}\Bigr\}\notag\\
&+\frac{1}{16\pi G_n}\int dt\int_{S_t}(-1)^{n-1}N^A\pi_BE^B_{\ A}-\Bigl\{2\hat{K}^A\land\phi_A\notag\\
&\qquad\qquad\qquad\qquad+\beta l^2\hat{K}^A\land\Bigl[\hat{\Omega}^{BC}-\hat{H}^B\land\hat{H}^C-\frac{1}{3}\hat{K}^B\land\hat{K}^C\Bigr]\land\phi_{ABC}\Bigr\}\notag\\
&-\frac{1}{16\pi G_n}\int dt\int_{\partial S_t}\beta l^2N\Bigl\{b_CE^Cl^{AB}+2a^A\hat{K}^B\Bigr\}\land\phi_{AB}.
\end{align}

\section{deSitter entropy coincides with hologra-phic entanglement entropy}
\label{sec:entropy}

In this section, we derive the de Sitter entropy for general stationary asymptotically de Sitter braneworld spacetimes in the RS II model using the canonical formalism introduced in the previous section and compare it with the holographic entanglement entropy. We show that, for general stationary spacetimes, the two entropies do not coincide because the extrinsic curvature of constant-time hypersurfaces does not vanish. In contrast, for static spacetimes, the extrinsic curvature of constant-time hypersurfaces vanishes, and consequently the de Sitter entropy coincides with the holographic entanglement entropy.

The holographic entanglement entropy with Gauss-Bonnet is given by
\begin{align}
S_{\rm JM}=\frac{1}{4G_n}\int_{\Gamma^+\cup\Gamma^-}d^{n-2}x\sqrt{G}\left(1+\frac{\beta l^2}{2}\ ^{(n-2)}\tilde{R}\right)+\frac{1}{2G_n}\int_{\partial\Gamma}d^{n-3}x\sqrt{p}\frac{\beta l^2}{2}[\ ^{(n-3)}K]^-,
\end{align}
where $G_{xy}$ is the induced metricof $(n-2)$-dimensional surface $\Gamma^\pm$ with $\partial\Gamma^+=\partial\Gamma^-=:\partial\Gamma$, $^{(n-2)}\tilde{R}$ is the Ricci scalar of $G_{xy}$, $p_{AB}$ is the induced metric of $\partial\Gamma$ and $^{(n-3)}K$ is the extrinsic curvature of $\partial\Gamma$. We suppose $\Gamma^+\subset M_n^+,\Gamma^-\subset M_n^-$ and $\partial\Gamma\subset M_{n-1}$ in the braneworld setup.

We will show that the de Sitter entropy coincides with the holographic entanglement entropy in Gauss-Bonnet gravity. Since the de Sitter entropy is obtained from the Euclidean action, we consider the $n$-dimensional stationary spacetime $M_n^+ \cup M_n^-$ and Euclideanize $M_n^+ \cup M_n^-$. We choose the time coordinate such that, on the horizon, the corresponding time-evolution vector is proportional to the null generator of the horizon. This coordinate choice does not restrict the generality of our analysis. Moreover, we assume that the imaginary time $\tau = it$ in the Euclideanized spacetime is periodic with period $\beta_0$. Note that the generator of the cosmological horizon $\Gamma^+ \cup \Gamma^-$ is a Killing vector. Therefore, the extrinsic curvature of $\Gamma^+ \cup \Gamma^-$ vanishes. In this paper, the subscript $E$ denotes the Euclideanized spacetime. The metric of the Euclideanized spacetime is given by
\begin{align}
ds_E&=N^2d\tau^2+h_{ij}[dx^i-iN^id\tau][dx^j-iN^jd\tau],
\end{align}
Furthermore, assuming that the location of the horizon
$\mathcal{H}_E:=(\Gamma^+_\epsilon\cup\Gamma^-_\epsilon)\times S^1$ on $\Sigma_E$ is characterized by $N(x)=0$, and that the spatial
shift can be chosen such that $N^a=0$ on $\mathcal{H}_E$, the metric near $\mathcal{H}_E$ can be written as
\begin{align}
ds_E^2\approx N^2d\tau^2+h_{ij}dx^idx^j,
\end{align}
as discussed in Ref.~\cite{Brown:1993ke}. Since $N(x)$ is independent of the imaginary time $\tau$, the orbit of $\tau$ on each $N=\mathrm{constant}$ surface in $M_{E,n}^+\cup M_{E,n}^-$ is homeomorphic to $S^1$. At $\mathcal{H}_E$, the $S^1$ orbit shrinks to zero size, resulting in a conical singularity. To regularize this singular behavior, we introduce a cutoff surface defined by $N(x)=\epsilon$ and consider the spacetime bounded by $M_{E,n-1}$ and $\mathcal{H}_{E,\epsilon}$, where $\mathcal{H}_{E,\epsilon}$ is defined by $N(x)=\epsilon$. In this work, the subscript $\epsilon$ denotes a spacetime with a cutoff surface.. After all calculations are completed, we take the limit $\epsilon\rightarrow0$. Note that $\mathcal{H}_{E,\epsilon}$ is a virtual boundary and no
boundary condition is imposed on it. Furthermore, the condition for the absence of a conical singularity is obtained by taking the limit
$\epsilon\rightarrow0$. The metric in the neighborhood of $\mathcal{H}_E$ is
\begin{align}
ds^2_E&\approx N^2d\tau^2+\alpha^2dN^2+\cdots \notag\\
&=\rho^2d\left(\frac{\tau}{\alpha}\right)^2+d\rho^2+\cdots,
\end{align}
where
\begin{align}
\alpha^2&=g_{NN}\big|_{\Gamma^+_\epsilon\cup\Gamma^-_\epsilon},
\end{align}
and we have introduced the radial coordinate $\rho=\alpha N$.

Since
\begin{align}
\alpha^{-1}&=\lim_{\epsilon\rightarrow0}r^M\partial_MN
\end{align}
on $\Gamma^+_\epsilon\cup\Gamma^-_\epsilon$, where $r^M$ is the unit normal vector to $(\Gamma^+_\epsilon\cup\Gamma^-_\epsilon)\times S^1$, the condition for the absence of a conical singularity is
\begin{align}
\label{conical condition}
\lim_{\epsilon\rightarrow0}\beta_0 r^M\partial_MN
&=2\pi,
\qquad \text{on }\Gamma^+_\epsilon\cup\Gamma^-_\epsilon,
\end{align}
as discussed in Refs.~\cite{Solodukhin:1994yz,Fursaev:1995ef}. Under this condition, the deficit angle vanishes, and no conical
singularity arises at $\mathcal{H}_E$.

Then, the regularied Euclidean action is
\begin{align} 
I_{E,\epsilon}&=\frac{1}{16\pi G_n}\int_{M_{n,E,\epsilon}^{+}\cup M_{n,E,\epsilon}^-}d^nx\sqrt{-g}\left(-R+2\Lambda-\frac{\beta l^2}{4}\mathcal{L}_{GB}\right)\notag\\
&\qquad\qquad-\int_{M_{n-1,E,\epsilon}}d^{n-1}x\sqrt{-q}\left(-\sigma+\frac{1}{16\pi G_n}[Q]^-\right).
\end{align}
We see that the deSitter entropy in the braneworld with the Gauss-Bonnetterm is given by
\begin{align}
\label{deSitter entropy}
S_{\rm dS}+J&=-\lim_{\epsilon\rightarrow0}I_{E,\epsilon}\notag\\
&=\lim_{\epsilon\rightarrow0}\Biggl[\frac{1}{16\pi G_n}\int d\tau\int_{\Sigma_{n,E,\epsilon}}d^{n-1}x\big(i\pi^{ij}\dot{h}_{ij}-N\mathcal{H}-N^i\mathcal{H}_{i}\big)\notag\\
&\ \ \ \ \ \ \ \ \ \frac{1}{16\pi G_n}\int d\tau\int_{\Gamma^+_{E,\epsilon}\cup\Gamma^-_{E,\epsilon}}(-1)^nN^i\pi_jh^j_{\ i}+N\Bigl\{2a^b\xi_b+\beta l^2\Bigl[-H^b\land\$_\perp\tilde{\omega}^{cd}+a^bF^{cd}+\tilde{\nabla}H^bl^{cd}\Bigr]\land\xi_{bcd}\Bigr\}\notag\\
&\ \ \ \ \ \ \ \ \ +\int d\tau\int_{\Sigma_{n-1,E,\epsilon}}d^{n-2}xN\sqrt{\gamma}\Big(-\sigma+\frac{1}{16\pi G_n}\Bigl[-\frac{1}{N\sqrt{h}}n^iN^j\pi_{ij}+2K+2K_{\mu\nu}u^{\mu}u^{\nu}\Bigr]^-\notag\\
&\qquad\qquad\qquad\qquad+\frac{\beta l^2}{16\pi G_n}\Bigl[\frac{1}{2}\gamma^{X_1Y_1X_2}_{\ Z_1W_1Z_2}K^{Z_1}_{\ \ X_1}{}^{(n-1)}R^{W_1Z_2}_{\ \ \ Y_1X_2}-\frac{1}{3}\gamma^{X_1Y_1X_2}_{\ Z_1W_1Z_2}K^{Z_1}_{\ \ X_1}K^{W_1}_{\ \ Y_1}K^{Z_1}_{\ \ X_2}\Bigr]^-\notag\\
&\ \ \ \ \ \ \ \ \ +\frac{\beta l^2}{16\pi G_n}\Bigl[\gamma^{X_1Y_1X_2}_{\ Z_1W_1Z_2}K^{Z_1}_{\ \ X_1}H^{W_1}_{\ \ Y_1}H^{Z_2}_{\ \ X_2}\Bigr]^-\Big)-\frac{1}{16\pi G_n}\int d\tau\int_{\partial\Gamma_{E,\epsilon}}\beta l^2N\Bigl\{b_CE^Cl^{AB}+2a^A\hat{K}^B\Bigr\}\land\phi_{AB}\Biggr]\notag\\
&=\lim_{\epsilon\rightarrow0}\Biggl[\frac{\beta_0}{16\pi G_n}\int_{\Gamma^+_{E,\epsilon}\cup\Gamma^-_{E,\epsilon}}(-1)^nN^i\pi_jh^j_{\ i}+N\Bigl\{2a^b\xi_b+\beta l^2\Bigl[-H^b\land\$_\perp\tilde{\omega}^{cd}+a^bF^{cd}+\tilde{\nabla}H^bl^{cd}\Bigr]\land\xi_{bcd}\Bigr\}\notag\\
&\ \ \ \ \ \ \ \ \ -\frac{\beta_0}{16\pi G_n}\int_{\partial\Gamma_{E,\epsilon}}\beta l^2N\Bigl\{b_CE^Cl^{AB}+2a^A\hat{K}^B\Bigr\}\land\phi_{AB}\Biggr]\notag\\
&\ \ \ \ \ \ \ \ \ +\beta_0\int_{\Sigma_{n-1,E}}d^{n-2}xN\sqrt{\gamma}\Big(-\sigma+\frac{1}{16\pi G_n}\Bigl[2K+2K_{\mu\nu}u^{\mu}u^{\nu}\Bigr]^-\notag\\
&\qquad\qquad\qquad\qquad+\frac{\beta l^2}{16\pi G_n}\Bigl[\frac{1}{2}\gamma^{X_1Y_1X_2}_{\ Z_1W_1Z_2}K^{Z_1}_{\ \ X_1}{}^{(n-1)}R^{W_1Z_2}_{\ \ \ Y_1X_2}-\frac{1}{3}\gamma^{X_1Y_1X_2}_{\ Z_1W_1Z_2}K^{Z_1}_{\ \ X_1}K^{W_1}_{\ \ Y_1}K^{Z_2}_{\ \ X_2})\Bigr]^-\notag\\
&\ \ \ \ \ \ \ \ \  +\frac{\beta l^2}{8\pi G_n}\Bigl[\gamma^{X_1Y_1X_2}_{\ Z_1W_1Z_2}K^{Z_1}_{\ \ X_1}H^{W_1}_{\ \ Y_1}H^{Z_2}_{\ \ X_2}\Bigr]^-\Big)+J,\\
J&:=\beta_0\int_{\Sigma_{n-1}}d^{n-2}x\sqrt{\gamma}\Bigl[-\frac{1}{\sqrt{h}}n^iN^j\pi_{ij}\Bigr]^-,
\end{align}
where $\mathcal{H}$ is the Hamiltonian constraint, $\mathcal{H}_i$ is the momentum constraint, $\Sigma_{E,\epsilon}$ is the imaginary time constant surface, $\Sigma_{n-1,\epsilon}$ is $\Sigma_{E,\epsilon}\cap M_{E,n-1,\epsilon}$, $J$ is the angular momentum, the tensor $\gamma^{X_1Y_1X_2}_{\ Z_1W_1Z_2}$ is defined as 
\begin{align}
\gamma^{X_1Y_1X_2}_{\ Z_1W_1Z_2}:=3!\gamma^{X_1}_{\ \ [Z_1}\gamma^{Y_1}_{\ \ W_1}\gamma^{X_2}_{\ \ Z_2]},
\end{align}
and, in the third equality of Eq. (\ref{deSitter entropy}), we used the facts that 
\begin{align}
\mathcal{H}=\mathcal{H}_M=\dot{h}_{ij}=0.
\end{align}
Also, in the second equality of Eq. (\ref{deSitter entropy}), we used the following relation equations for the last term on $\partial\Gamma$.
\begin{align}
&-\lim_{\epsilon\rightarrow0}\frac{\beta_0\beta l^2}{16\pi G_n}\int_{\partial\Gamma_{E,\epsilon}}\Bigl\{b_CE^Cl^{AB}+2a^A\hat{K}^B\Bigr\}\land\phi_{AB}\notag\\
=&\lim_{\epsilon\rightarrow0}\frac{\beta_0\beta l^2}{8\pi G_n}\int_{\partial\Gamma_{E,\epsilon}}d^{n-3}x\sqrt{p}Na^Mr_M{}^{(n-3)}K\notag\\
=&\frac{\beta l^2}{4G_n}\int_{\partial\Gamma_E}d^{n-3}x\sqrt{p}{}^{(n-3)}K,
\end{align}
where in the first equality, we used the Stokes theorem and the fact that $b_Z=H_{ZM}n^M$ and the extrinsic curvature $H_{ab}$ vanishes, and in the second equality, the final result is derived from the condition for the conical singularity not to occur (\ref{conical condition}) and the definition of $a^M$ in Eq.~(\ref{a}). 

The term of the deSitter entropy on $\Gamma^+_\epsilon\cup\Gamma^-_\epsilon$ is 
\begin{align}
&\lim_{\epsilon\rightarrow0}\frac{1}{16\pi G_n}\int dt\int_{\Gamma^+_{E,\epsilon}\cup\Gamma^-_{E,\epsilon}}(-1)^nN^i\pi_jh^j_{\ i}+N\Bigl\{2a^b\xi_b+\beta l^2\Bigl[-H^b\land\$_\perp\tilde{\omega}^{cd}+a^bF^{cd}+\tilde{\nabla}H^bl^{cd}\Bigr]\land\xi_{bcd}\Bigr\}\notag\\
=&\lim_{\epsilon\rightarrow0}\frac{\beta_0}{16\pi G_n}\int_{\Gamma^+_{E,\epsilon}\cup\Gamma^-_{E,\epsilon}}d^{n-2}x\sqrt{G}\left(2r^a\partial_aN+\beta l^2r^a\partial_aN\ ^{(n-2)}\tilde{R}\right)\notag\\
=&\frac{1}{4G_n}\int_{\Gamma^+_E\cup\Gamma^-_E}d^{n-2}x\sqrt{G}\left(1+\frac{\beta l^2}{2}\ ^{(n-2)}\tilde{R}\right),
\end{align}
where, in the first equality, we have used the fact that the extrinsic curvature $H_{ij}$ of the constant-time hypersurface vanishes on $\Gamma^+_{E,\epsilon}\cup\Gamma^-_{E,\epsilon}$, as does the shift vector. In the last equality, we have used the condition (\ref{conical condition}).

In the last step, we chack that the term of the deSitter entropy on $\Sigma_{n-1,E,\epsilon}$ vanishes using the junction condition (\ref{junction condition}). At first, we find the following equation from the junction condition.
\begin{align}
\label{junction condition 2}
&[K_{\mu\nu}u^{\mu}u^{\nu}+K]^-+\frac{\beta l^2}{2}[3J_{\mu\nu}u^{\mu}u^{\nu}+J-2P_{\mu\rho\nu\sigma}u^{\mu}u^{\nu}K^{\rho\sigma}]^-=8\pi G_n\sigma,
\end{align}
where 
\begin{align}
&J_{\mu\nu}u^\mu u^\nu=-2K_{\mu\rho}u^\mu K_{\nu\sigma}u^\nu K^{\rho\sigma}+2KK_{\mu\rho}u^\mu K^\rho_{\ \ \nu}u^\nu+K_{\mu\nu}u^\mu u^\nu K_{\rho\sigma}K^{\rho\sigma}-K^2K_{\mu\nu}u^\mu u^\nu,\\
&P_{\mu\rho\nu\sigma}u^{\mu}u^{\nu}K^{\rho\sigma}={}^{(n-1)}R_{\mu\rho\nu\sigma}u^{\mu}u^{\nu}K^{\rho\sigma}-{}^{(n-1)}R_{\mu\nu}u^\mu u^\nu K+2{}^{(n-1)}R_{\mu\rho}u^\mu K^{\rho}_{\ \nu}u^\nu\notag\\
&\qquad\qquad\qquad\qquad+{}^{(n-1)}G_{\mu\nu}K^{\mu\nu}-\frac{1}{2}{}^{(n-1)}RK_{\mu\nu}u^\mu u^\nu,
\end{align}
The term of the deSitter entropy on $\Sigma_{n-1,E,\epsilon}$ Eq. (\ref{deSitter entropy}) is
\begin{align}
&\beta_0\int_{\Sigma_{n-1,E,\epsilon}}d^{n-2}xN\sqrt{\gamma}\Big(-\sigma+\frac{1}{8\pi G_n}\Bigl[K_{\mu\nu}u^{\mu}u^{\nu}+K\Bigr]^-\notag\\
&\qquad\qquad\qquad\qquad+\beta l^2\Bigl[\frac{1}{2}\gamma^{X_1Y_1X_2}_{\ Z_1W_1Z_2}K^{Z_1}_{\ \ X_1}{}^{(n-1)}R^{W_1Z_2}_{\ \ \ Y_1X_2}-\frac{1}{3}\gamma^{X_1Y_1X_2}_{\ Z_1W_1Z_2}K^{Z_1}_{\ \ X_1}K^{W_1}_{\ \ Y_1}K^{Z_2}_{\ \ X_2}\Bigr]^-\notag\\
&\qquad\qquad\qquad\qquad-2\beta l^2\Bigl[\gamma^{X_1Y_1X_2}_{\ Z_1W_1Z_2}K^{Z_1}_{\ \ X_1}H^{W_1}_{\ \ Y_1}H^{Z_2}_{\ \ X_2}\Bigr]^-\Big).
\end{align}
Then, we see that the first terms of the second line of Eq. (\ref{deSitter entropy}) can be expressed as
\begin{align}
&\frac{1}{2}\gamma^{X_1Y_1X_2}_{\ Z_1W_1Z_2}K^{Z_1}_{\ \ X_1}{}^{(n-1)}R^{W_1Z_2}_{\ \ \ Y_1X_2}\notag\\
&=\frac{1}{2}q^{\mu_1\nu_1\mu_2}_{\ \rho_1\sigma_1\rho_2}K^{\rho_1}_{\ \ \mu_1}{}^{(n-1)}R^{\sigma_1\rho_2}_{\ \ \ \nu_1\mu_2}+\frac{1}{2}3!\ u^{\mu_1}u_{[\rho_1}q^{\nu_1}_{\ \sigma_1}q^{\mu_2}_{\ \rho_2]}K^{\rho_1}_{\ \ \mu_1}{}^{(n-1)}R^{\sigma_1\rho_2}_{\ \ \ \nu_1\mu_2}\notag\\
&\ \ \ +3!\ q^{\mu_1}_{\ [\rho_1}u^{\nu_1}u_{\sigma_1}q^{\mu_2}_{\ \rho_2]}K^{\rho_1}_{\ \ \mu_1}{}^{(n-1)}R^{\sigma_1\rho_2}_{\ \ \ \nu_1\mu_2}\notag\\
&=-2{}^{(n-1)}R_{\mu\rho\nu\sigma}u^{\mu}u^{\nu}K^{\rho\sigma}+2{}^{(n-1)}R_{\mu\nu}u^\mu u^\nu K-4{}^{(n-1)}R_{\mu\rho}u^\mu K^{\rho}_{\ \nu}u^\nu\notag\\
&\ \ \ \ -2{}^{(n-1)}G_{\mu\nu}K^{\mu\nu}+{}^{(n-1)}RK_{\mu\nu}u^\mu u^\nu\notag\\
&=-2P_{\mu\rho\nu\sigma}u^\mu u^\nu K^{\rho\sigma},
\end{align}
where in the second equality, we used the following equations.
\begin{align}
&\frac{1}{2}q^{\mu_1\nu_1\mu_2}_{\ \rho_1\sigma_1\rho_2}K^{\rho_1}_{\ \ \mu_1}{}^{(n-1)}R^{\sigma_1\rho_2}_{\ \ \ \nu_1\mu_2}\notag\\
&=-2{}^{(n-1)}G_{\mu\nu}K^{\mu\nu},\\
&\frac{3!}{2}u^{\mu_1}u_{[\rho_1}q^{\nu_1}_{\ \sigma_1}q^{\mu_2}_{\ \rho_2]}K^{\rho_1}_{\ \ \mu_1}{}^{(n-1)}R^{\sigma_1\rho_2}_{\ \ \ \nu_1\mu_2}\notag\\
&={}^{(n-1)}RK_{\mu\nu}u^\mu u^\nu-2{}^{(n-1)}R_{\mu\rho}u^{\mu}K^{\rho}_{\ \nu}u^{\nu},\\
&3!\ q^{\mu_1}_{\ [\rho_1}u^{\nu_1}u_{\sigma_1}q^{\mu_2}_{\ \rho_2]}K^{\rho_1}_{\ \ \mu_1}{}^{(n-1)}R^{\sigma_1\rho_2}_{\ \ \ \nu_1\mu_2}\notag\\
&=-2{}^{(n-1)}R_{\mu\rho\nu\sigma}u^\mu u^\nu K^{\rho\sigma}+2{}^{(n-1)}R_{\mu\nu}u^\mu u^\nu K-2{}^{(n-1)}R^{\rho}_{\ \mu}u^\mu K^{\rho}_{\ \nu}u^\nu.
\end{align}
Also, we find that the second term of the second line of Eq. (\ref{deSitter entropy}) is computed as
\begin{align}
&-\frac{1}{3}\gamma^{X_1Y_1X_2}_{\ Z_1W_1Z_2}K^{Z_1}_{\ \ X_1}K^{W_1}_{\ \ Y_1}K^{Z_2}_{\ \ X_2}\notag\\
&=-3!u^{\mu_1}u_{[\rho_1}\gamma^{\nu_1}_{\ \sigma_1}\gamma^{\mu_2}_{\ \rho_2]}K^{\rho_1}_{\ \ \mu_1}K^{\sigma_1}_{\ \ \nu_1}K^{\rho_2}_{\ \ \mu_2}-\frac{1}{3}q^{\mu_1\nu_1\mu_2}_{\ \rho_1\sigma_1\rho_2}K^{\rho_1}_{\ \ \mu_1}K^{\sigma_1}_{\ \ \nu_1}K^{\rho_2}_{\ \ \mu_2}\notag\\
&=3J_{\mu\nu}u^\mu u^\nu+J,
\end{align}
where in the second equality, we used the fact following as
\begin{align}
&-3!u^{\mu_1}u_{[\rho_1}\gamma^{\nu_1}_{\ \sigma_1}\gamma^{\mu_2}_{\ \rho_2]}K^{\rho_1}_{\ \ \mu_1}K^{\sigma_1}_{\ \ \nu_1}K^{\rho_2}_{\ \ \mu_2}\notag\\
&=-3!u^{\mu_1}u_{[\rho_1}q^{\nu_1}_{\ \sigma_1}q^{\mu_2}_{\ \rho_2]}K^{\rho_1}_{\ \ \mu_1}K^{\sigma_1}_{\ \ \nu_1}K^{\rho_2}_{\ \ \mu_2}\notag\\
&=-2K_{\mu\rho}u^\mu K_{\nu\sigma}u^\nu K^{\rho\sigma}+2KK_{\mu\rho}u^\mu K^\rho_{\ \ \nu}u^\nu+K_{\mu\nu}u^\mu u^\nu K_{\rho\sigma}K^{\rho\sigma}-K^2K_{\mu\nu}u^\mu u^\nu\notag\\
&=3J_{\mu\nu}u^\mu u^\nu.
\end{align}
Therefore, the term of Eq. (\ref{deSitter entropy}) on $\Sigma_{n-1,E,\epsilon}$ becomes
\begin{align}
&\beta_0\int_{\Sigma_{n-1,E,\epsilon}}d^{n-2}xN\sqrt{\gamma}\Big(-\sigma+\frac{1}{8\pi G_n}\Bigl[K_{\mu\nu}u^{\mu}u^{\nu}+K\Bigr]^-+\frac{\beta l^2}{16\pi G_n}[3J_{\mu\nu}u^{\mu}u^{\nu}+J-2P_{\mu\rho\nu\sigma}u^{\mu}u^{\nu}K^{\rho\sigma}]^-\notag\\
&\qquad\qquad +\frac{\beta l^2}{8\pi G_n}\Bigl[\gamma^{X_1Y_1X_2}_{\ Z_1W_1Z_2}K^{Z_1}_{\ \ X_1}H^{W_1}_{\ \ Y_1}H^{Z_2}_{\ \ X_2}\Bigr]^-\Big)\notag\\
&=\frac{\beta_0}{8\pi G_n}\int_{\Sigma_{n-1,E,\epsilon}}d^{n-2}xN\sqrt{\gamma}\beta l^2\Bigl[\gamma^{X_1Y_1X_2}_{\ Z_1W_1Z_2}K^{Z_1}_{\ \ X_1}H^{W_1}_{\ \ Y_1}H^{Z_2}_{\ \ X_2}\Bigr]^-,
\end{align}
where we used the junction condition Eq. (\ref{junction condition 2}).\\

Therefore, the de Sitter entropy is 
\begin{align}
S_{{\rm dS}}=&\frac{1}{4G_n}\int_{\Gamma^+_E\cup\Gamma^-_E}d^{n-2}x\sqrt{G}\left(1+\frac{\beta l^2}{2}\ ^{(n-2)}\tilde{R}\right)+\frac{1}{2G_n}\int_{\partial\Gamma_E}d^{n-3}x\sqrt{p}\frac{\beta l^2}{2}[\ ^{(n-3)}K]^-\notag\\
&+\frac{\beta_0}{8\pi G_n}\int_{\Sigma_{n-1,E}}d^{n-2}xN\sqrt{\gamma}\beta l^2\Bigl[\gamma^{X_1Y_1X_2}_{\ Z_1W_1Z_2}K^{Z_1}_{\ \ X_1}H^{W_1}_{\ \ Y_1}H^{Z_2}_{\ \ X_2}\Bigr]^-.
\end{align}

As a result, we can find that 
\begin{align}
S_{{\rm dS}}\neq S_{{\rm JM}},
\end{align}
do not coincide due to the presence of the term involving the extrinsic curvature $H_{ab}$ of the boundary. Therefore, the deSitter entropy corresponds to the holographic entanglement entropy in the static spacetime but not in the stationary spacetime.

\section{Agreement between de Sitter Entropy and Holographic Entanglement Entropy in Static Lovelock Gravity}
\label{sec:lovelock}

In the previous section, we showed that, in Gauss-Bonnet gravity, the de Sitter entropy and the holographic entanglement entropy generally do not coincide for stationary asymptotically de Sitter braneworld spacetimes in the RS II model, whereas they coincide for static spacetimes in which the extrinsic curvature of the constant-time hypersurfaces vanishes. In this section, we extend the analysis to general Lovelock gravity and show that the de Sitter entropy coincides with the holographic entanglement entropy for general static asymptotically de Sitter braneworld spacetimes. The Lovelock action with boundary is given in Ref~\cite{Myers:1987yn,Maeda:2003vq,Cano:2018ckq},
\begin{align}
I=\frac{1}{16\pi G_n}\int_{M^+_n\cup M^-_n}d^nx\sqrt{-g}(-2\Lambda+\sum_{m}c_m\mathcal{L}_m)+\int_{M_{n-1}}d^{n-1}x\sqrt{-q}\left(-\sigma+\sum_mc_m\frac{[Q_m]^-}{16\pi G_n}\right),
\label{action lovelock}
\end{align}
where $c_m$ are coeffcients of Lovelock terms, 
\begin{align}
\mathcal{L}_m&=\frac{1}{2^m}g^{M_1N_1M_2N_2\cdots M_mN_m}_{L_1K_1L_2K_2\cdots L_mK_m}R_{M_1N_1}^{\ \ \ L_1K_1}R_{M_2N_2}^{\ \ \ L_2K_2}\cdots R_{M_mN_m}^{\ \ \ L_mL_m},\\
Q_m&=\frac{4m}{2^m}\int^1_0ds\ q^{\mu_1\nu_1\mu_2\nu_2\cdots \mu_{m-1}\nu_{m-1}\mu_m}_{\rho_1\sigma_1\rho_2d\sigma_2\cdots \rho_{m-1}\sigma_{m-1}\rho_m}({}^{(n-1)}R_{\mu_1\nu_1}^{\ \ \ \rho_1\sigma_1}-2s^2K^{\rho_1}_{\ \ \mu_1}K^{\sigma_1}_{\ \ \sigma_1})({}^{(n-1)}R_{\mu_2\nu_2}^{\ \ \ \rho_2\sigma_2}-2s^2K^{\rho_2}_{\ \ \mu_2}K^{\sigma_2}_{\ \ \nu_2})\notag\\
&\qquad\qquad\qquad\qquad\qquad\qquad\qquad\cdots({}^{(n-1)}R_{\mu_{m-1}\nu_{m-1}}^{\ \ \ \rho_{m-1}\sigma_{m-1}}-2s^2K^{\rho_{m-1}}_{\ \ \mu_{m-1}}K^{\sigma_{m-1}}_{\ \ \nu_{m-1}})K^{\rho_m}_{\ \ \mu_m}\notag\\
&=\frac{4m}{2^m}q^{\mu_1\nu_1\mu_2\nu_2\cdots \mu_{m-1}\nu_{m-1}\mu_m}_{\rho_1\sigma_1\rho_2d\sigma_2\cdots \rho_{m-1}\sigma_{m-1}\rho_m}\sum_{k=0}^{m-1}\binom{n-1}{k}\frac{(-2)^k}{2k+1}K^{\rho_1}_{\ \ \mu_1}K^{\sigma_1}_{\ \ \sigma_1}K^{\rho_2}_{\ \ \mu_2}K^{\sigma_2}_{\ \ \nu_2}\cdots K^{\rho_k}_{\ \ \mu_k}K^{\sigma_k}_{\ \ \nu_k}K^{\rho_m}_{\ \ \mu_m}\notag\\
&\qquad\qquad\qquad\qquad\qquad\qquad\qquad\qquad\qquad\times{}^{(n-1)}R_{\mu_{k+1}\nu_{k+1}}^{\ \ \ \rho_{k+1}\sigma_{k+1}}\cdots{}^{(n-1)}R_{\mu_{m-1}\nu_{m-1}}^{\ \ \ \rho_{m-1}\sigma_{m-1}},
\end{align}
and $\binom{m-1}{k}$ is the binomial coefficients.The tensros $g^{M_1N_1M_2N_2\cdots M_mN_m}_{L_1K_1L_2K_2\cdots L_mK_m}$ and $q^{\mu_1\nu_1\mu_2\nu_2\cdots \mu_{m-1}\nu_{m-1}\mu_m}_{\rho_1\sigma_1\rho_2d\sigma_2\cdots \rho_{m-1}\sigma_{m-1}\rho_m}$ are defined as
\begin{align}
g^{M_1N_1M_2N_2\cdots M_mN_m}_{L_1K_1L_2K_2\cdots L_mK_m}&:=(2m)!\ \delta^{M_1}_{\ \ [L_1}\delta^{N_1}_{\ \ K_1}\cdots\delta^{M_m}_{\ \ L_m}\delta^{N_m}_{\ \ K_m]},\\
q^{\mu_1\nu_1\mu_2\nu_2\cdots \mu_{m-1}\nu_{m-1}\mu_m}_{\rho_1\sigma_1\rho_2d\sigma_2\cdots \rho_{m-1}\sigma_{m-1}\rho_m}&:=(2m-1)!\ q^{\mu_1}_{\ \ [\rho_1}q^{\nu_1}_{\ \ \sigma_1}\cdots q^{\mu_{m-1}}_{\ \ \rho_{m-1}}q^{\nu_{m-1}}_{\ \ \sigma_{m-1}}q^{\mu_m}_{\ \ \rho_{m}]}.
\end{align}
Here, the brackets $[\ ]$ denote complete antisymmetrization over the enclosed indices.

The junction condition is 
\begin{align}
\sigma q_{\mu\nu}+\frac{1}{16\pi G_n}\sum_m c_m[J_{\mu\nu}^m]^-=0,
\end{align}
where
\begin{align}
J_{\mu\nu}^m=&\frac{4m}{2^m}q^{\mu_1\nu_1\mu_2\nu_2\cdots \mu_{m-1}\nu_{m-1}\mu_m}_{\rho_1\sigma_1\rho_2d\sigma_2\cdots \rho_{m-1}\sigma_{m-1}\rho_m}\sum_{k=0}^{m-1}\binom{m-1}{k}\frac{(-2)^k}{2k+1}\Bigl((2k+1)K^{\rho_1}_{\ \ \mu_1}K^{\sigma_1}_{\ \ \nu_1}\cdots K^{\rho_k}_{\ \ \mu_k}K^{\sigma_k}_{\ \ \nu_k}K^{\rho_m}_{\ \ \mu}q_{\mu_m\nu}\notag\\
&\times{}^{(n-1)}R_{\mu_{k+1}\nu_{k+1}}^{\ \ \ \rho_{k+1}\sigma_{k+1}}\cdots{}^{(n-1)}R_{\mu_{m-1}\nu_{m-1}}^{\ \ \ \rho_{m-1}\sigma_{m-1}}+2(m-k-1)K^{\rho_1}_{\ \ \mu_1}K^{\sigma_1}_{\ \ \nu_1}\cdots K^{\rho_k}_{\ \ \mu_k}K^{\sigma_k}_{\ \ \nu_k}K^{\rho_m}_{\ \ \mu_m}\notag\\
&\times{}^{(n-1)}R_{\mu_{k+1}\nu_{k+1}}^{\ \ \ \rho_{k+1}\sigma_{k+1}}\cdots{}^{(n-1)}R_{\mu_{m-2}\nu_{m-2}}^{\ \ \ \rho_{m-2}\sigma_{m-2}}{}^{(n-1)}R_{\mu_{m-1}\nu_{m-1}}^{\ \ \ \rho_{m-1}\rho}q_{\rho\mu}q^{\sigma_{m-1}}_{\ \ \nu}\Bigr)-Q_mq_{\mu\nu}.
\end{align}
Contracting the tensor with $u^\mu$ on both indices, we obtain
\begin{align}
&-\sigma+\frac{1}{16\pi G_n}\sum_mc_m\Biggl[\frac{4m}{2^m}q^{\mu_1\nu_1\mu_2\nu_2\cdots \mu_{m-1}\nu_{m-1}\mu_m}_{\rho_1\sigma_1\rho_2d\sigma_2\cdots \rho_{m-1}\sigma_{m-1}\rho_m}\sum_{k=0}^{m-1}\binom{m-1}{k}\frac{(-2)^k}{2k+1}\Bigl((2k+1)K^{\rho_1}_{\ \ \mu_1}K^{\sigma_1}_{\ \ \nu_1}\cdots K^{\rho_k}_{\ \ \mu_k}K^{\sigma_k}_{\ \ \nu_k}K^{\rho_m}_{\ \ \mu}u^\mu u_{\mu_m}\notag\\
&\times{}^{(n-1)}R_{\mu_{k+1}\nu_{k+1}}^{\ \ \ \rho_{k+1}\sigma_{k+1}}\cdots{}^{(n-1)}R_{\mu_{m-1}\nu_{m-1}}^{\ \ \ \rho_{m-1}\sigma_{m-1}}+2(m-k-1)K^{\rho_1}_{\ \ \mu_1}K^{\sigma_1}_{\ \ \nu_1}\cdots K^{\rho_k}_{\ \ \mu_k}K^{\sigma_k}_{\ \ \nu_k}K^{\rho_m}_{\ \ \mu_m}\notag\\
&\times{}^{(n-1)}R_{\mu_{k+1}\nu_{k+1}}^{\ \ \ \rho_{k+1}\sigma_{k+1}}\cdots{}^{(n-1)}R_{\mu_{m-2}\nu_{m-2}}^{\ \ \ \rho_{m-2}\sigma_{m-2}}{}^{(n-1)}R_{\mu_{m-1}\nu_{m-1}}^{\ \ \ \rho_{m-1}\mu}u_{\mu}u^{\sigma_{m-1}}\Bigr)+Q_m\Biggr]^-\notag\\
=&-\sigma+\frac{1}{16\pi G_n}\sum_mc_m\Biggl[\frac{4m}{2^m}\sum_{k=0}^{m-1}\binom{m-1}{k}\frac{(-2)^k}{2k+1}\Bigl((2k+1)\gamma^{X_1Y_1X_2Y_2\cdots X_{m-1}Y_{m-1}}_{Z_1W_1Z_2W_2\cdots Z_{m-1}W_{m-1}}K_{\mu\nu}u^\mu u^\nu K^{Z_1}_{\ \ X_1}K^{W_1}_{\ \ Y_1}\cdots K^{Z_k}_{\ \ X_k}K^{W_k}_{\ \ Y_k}\notag\\
&\times{}^{(n-1)}R_{X_{k+1}Y_{k+1}}^{\ \ \ Z_{k+1}W_{k+1}}\cdots{}^{(n-1)}R_{X_{m-1}Y_{m-1}}^{\ \ \ Z_{m-1}W_{m-1}}+2(m-k-1)q^{\mu_1\nu_1\mu_2\nu_2\cdots \mu_{m-1}\nu_{m-1}\mu_m}_{\rho_1\sigma_1\rho_2d\sigma_2\cdots \rho_{m-1}\sigma_{m-1}\rho_m}K^{\rho_1}_{\ \ \mu_1}K^{\sigma_1}_{\ \ \nu_1}\cdots K^{\rho_k}_{\ \ \mu_k}K^{\sigma_k}_{\ \ \nu_k}K^{\rho_m}_{\ \ \mu_m}\notag\\
&\times{}^{(n-1)}R_{\mu_{k+1}\nu_{k+1}}^{\ \ \ \rho_{k+1}\sigma_{k+1}}\cdots{}^{(n-1)}R_{\mu_{m-2}\nu_{m-2}}^{\ \ \ \rho_{m-2}\sigma_{m-2}}{}^{(n-1)}R_{\mu_{m-1}\nu_{m-1}}^{\ \ \ \rho_{m-1}\mu}u_{\mu}u^{\sigma_{m-1}}\Bigr)+Q_m\Biggr]^-.
\end{align}

We will show that the de Sitter entropy obtained from the Euclidean action of Lovelock gravity coincides with the following holographic entanglement entropy for general static asymptotically de Sitter braneworld spacetimes.
\begin{align}
S_{\rm JM}=&\ \frac{1}{4G_n}\int_{\Gamma^+\cup\Gamma^-}d^{n-2}x\sqrt{G}\sum_m\frac{m}{2^{m-1}}c_mG^{x_1y_1\cdots x_{m-1}y_{m-1}}_{z_1w_1\cdots z_{m-1}w_{m-1}}{}^{(n-2)}\tilde{R}_{x_1y_1}^{\ \ \ z_1w_1}\cdots {}^{(n-2)}\tilde{R}_{x_{m-1}y_{m-1}}^{\ \ \ z_{m-1}w_{m-1}}\notag\\
&+\frac{1}{4G_n}\int_{\partial\Gamma}dx^{n-3}xN\sqrt{p}\Biggl[\sum_m\frac{4m(m-1)}{2^{m-1}}c_mp^{A_1B_1\cdots A_{m-1}}_{C_1D_1\cdots C_{m-1}}\sum_{k=0}^{m-2}\binom{m-2}{k}\frac{(-2)^k}{2k+1}{}^{(n-3)}K^{C_1}_{\ \ A_1}{}^{(n-3)}K^{D_1}_{\ \ B_1}\notag\\
&\cdots {}^{(n-3)}K^{C_{k}}_{\ \ A_{k}}{}^{(n-3)}K^{D_{k}}_{\ \ B_{k}}{}^{(n-3)}R_{A_{k+1}B_{k+1}}^{\ \ \ C_{k+1}D_{k+1}}\cdots {}^{(n-3)}R_{A_{m-2}B_{m-2}}^{\ \ \ C_{m-2}D_{m-2}}{}^{(n-3)}K^{C_{m-1}}_{\ \ A_{m-1}}\Biggr]^-,
\end{align}
where the tensors $G^{x_1y_1\cdots x_{m-1}y_{m-1}}_{z_1w_1\cdots z_{m-1}w_{m-1}}$ and $p^{A_1B_1\cdots A_{m-1}}_{C_1D_1\cdots C_{m-1}}$ are defined as
\begin{align}
G^{x_1y_1\cdots x_{m-1}y_{m-1}}_{z_1w_1\cdots z_{m-1}w_{m-1}}&:=(2m-2)!\ G^{x_1}_{\ \ [z_1}G^{y_1}_{\ \ w_1}\cdots G^{x_{m-1}}_{\ \ z_{m-1}}G^{y_{m-1}}_{\ \ w_{m-1}]},\\
p^{A_1B_1\cdots A_{m-1}}_{C_1D_1\cdots C_{m-1}}&:=(2m-3)!\ p^{A_1}_{\ \ [C_1}p^{B_1}_{\ \ D_1}\cdots p^{B_{m-2}}_{\ \ W_{m-2}}p^{A_{m-1}}_{\ \ C_{m-1}]}.
\end{align}

First, the $\mathcal{L}_m$ of bulk action become
\begin{align}
\mathcal{L}_m&=\frac{1}{2^m}h^{i_1j_1i_2j_2\cdots i_mj_m}_{k_1l_1k_2l_2\cdots k_ml_m}R_{i_1j_1}^{\ \ \ k_1l_1}R_{i_2j_2}^{\ \ \ k_2l_2}\cdots R_{i_mj_m}^{\ \ \ k_ml_m}\notag\\
&\qquad-\frac{2m}{2^m}(2m)!\ u^{M_1}u_{[K_1}h^{N_1}_{\ \ L_1}\cdots h^{M_m}_{\ \ K_m}h^{N_m}_{\ \ L_m]}R_{M_1N_1}^{\ \ \ K_1L_1}R_{M_2N_2}^{\ \ \ K_2L_2}\cdots R_{M_mN_m}^{\ \ \ K_mL_m}\notag\\
&=\frac{1}{2^m}h^{i_1j_1i_2j_2\cdots i_mj_m}_{k_1l_1k_2l_2\cdots k_ml_m}\tilde{R}_{i_1j_1}^{\ \ \ k_1l_1}\tilde{R}_{i_2j_2}^{\ \ \ k_2l_2}\cdots \tilde{R}_{i_mj_m}^{\ \ \ k_ml_m}\notag\\
&\qquad-\frac{4m}{2^m}(2m-1)!\ h^{i_1}_{\ \ [l_1}\cdots h^{i_m}_{\ \ k_m}h^{j_m}_{\ \ l_m]}R_{Mj_1}^{\ \ \ Nl_1}u^{M}u_{N}R_{i_2j_2}^{\ \ \ k_2l_2}\cdots R_{i_mj_m}^{\ \ \ k_ml_m}\notag\\
&=\frac{1}{2^m}h^{i_1j_1i_2j_2\cdots i_mj_m}_{k_1l_1k_2l_2\cdots k_ml_m}\tilde{R}_{i_1j_1}^{\ \ \ k_1l_1}\tilde{R}_{i_2j_2}^{\ \ \ k_2l_2}\cdots \tilde{R}_{i_mj_m}^{\ \ \ k_ml_m}\notag\\
&\qquad-\frac{4m}{2^m}(2m-1)!\ h^{j_1}_{\ \ [l_1}\cdots h^{i_m}_{\ \ k_m}h^{j_m}_{\ \ l_m]}h^{M}_{\ \ j_1}h^{N l_1}\nabla_{K}(\delta^{K}_{\ \ M}u^L\nabla_L u_N-u^K\nabla_{M}u_N)R_{i_2j_2}^{\ \ \ k_2l_2}\cdots R_{i_mj_m}^{\ \ \ k_ml_m}\notag\\
&=\frac{1}{2^m}h^{i_1j_1i_2j_2\cdots i_mj_m}_{k_1l_1k_2l_2\cdots k_ml_m}\tilde{R}_{i_1j_1}^{\ \ \ k_1l_1}\tilde{R}_{i_2j_2}^{\ \ \ k_2l_2}\cdots \tilde{R}_{i_mj_m}^{\ \ \ k_ml_m}\notag\\
&\qquad+\frac{4m}{2^m}(2m-1)!\ (\delta^{K}_{\ \ M}u^L\nabla_L u_\nu-u^K\nabla_{M}u_N)\nabla_{K}\left(h^{j_1}_{\ \ [l_1}\cdots h^{i_m}_{\ \ k_m}h^{j_m}_{\ \ l_m]}h^{M}_{\ \ j_1}h^{N l_1}R_{M_2N_2}^{\ \ \ K_2L_2}\cdots R_{M_mN_m}^{\ \ \ K_mL_m}\right)\notag\\
&\qquad-\nabla_{K}\left(\frac{4m}{2^m}(2m-1)!\ h^{b_1}_{\ \ [l_1}\cdots h^{i_m}_{\ \ k_m}h^{j_m}_{\ \ l_m]}h^{M}_{\ \ j_1}h^{N l_1}(\delta^{K}_{\ \ N}u^L\nabla_L u_N-u^K\nabla_{M}u_N)R_{i_2j_2}^{\ \ \ k_2l_2}\cdots R_{i_mj_m}^{\ \ \ k_ml_m}\right),
\end{align}
where the tensor $h^{i_1j_1i_2j_2\cdots i_mj_m}_{k_1l_1k_2l_2\cdots k_ml_m}$ is defined as
\begin{align}
h^{i_1j_1i_2j_2\cdots i_mj_m}_{k_1l_1k_2l_2\cdots k_ml_m}=2m!\ h^{i_i}_{\ \ [k_1}h^{j_1}_{\ \ l_1}\cdots h^{i_m}_{\ \ k_m}h^{j_m}_{\ \ j_m]},
\end{align}
the first term is the Hamiltonian constraint $\mathcal{H}$, in the first equality, we decompose the expression into the components with and without the time-direction components; in the second equality, we use the fact that the extrinsic curvature of the constant-time hypersurfaces vanishes for static spacetimes to rewrite the spacetime Riemann tensor in terms of the Riemann tensor of the induced metric on the constant-time hypersurfaces; in the third equality, we use the following relation;
\begin{align}
R_{MiNj}u^{M}u^{N}=h^{M}_{\ \ i}h^{N}_{\ \ j}\nabla_{K}(\delta^{K}_{\ \ M}u^L\nabla_L u_N-u^K\nabla_{M}u_N),
\end{align}
and in the fourth equality, we use the Leibniz rule. Furthermore, the second term can be shown to vanish by the following manipulations.
\begin{align}
&(\delta^{L}_{\ \ M}u^K\nabla_K u_N-u^L\nabla_{M}u_N)\nabla_{L}\left(h^{j_1}_{\ \ [l_1}\cdots h^{i_m}_{\ \ k_m}h^{j_m}_{\ \ l_m]}h^{M}_{\ \ j_1}h^{Nl_1}R_{i_2j_2}^{\ \ \ k_2l_2}\cdots R_{i_mj_m}^{\ \ \ k_ml_m}\right)\notag\\
=&(\delta^{L}_{\ \ M}u^K\nabla_K u_N-u^L\nabla_{M}u_N)\Bigl(\nabla_{L}(h^{j_1}_{\ \ [l_1}\cdots h^{i_m}_{\ \ k_m}h^{j_m}_{\ \ l_m]})h^{M}_{\ \ j_1}h^{N l_1}R_{i_2j_2}^{\ \ \ k_2l_2}\cdots R_{i_mj_m}^{\ \ \ k_ml_m}\notag\\
&+h^{j_1}_{\ \ [l_1}\cdots h^{i_m}_{\ \ k_m}h^{j_m}_{\ \ l_m]}\nabla_L(h^{\mu}_{\ \ bj_1}h^{\nu l_1})R_{i_2j_2}^{\ \ \ k_2l_2}\cdots R_{i_mj_m}^{\ \ \ k_ml_m}\notag\\
&+h^{j_1}_{\ \ [l_1}\cdots h^{i_m}_{\ \ k_m}h^{j_m}_{\ \ l_m]}h^{\mu}_{\ \ j_1}h^{N l_1}\nabla_{L}(R_{i_2j_2}^{\ \ \ k_2l_2}\cdots R_{i_mj_m}^{\ \ \ k_ml_m})\Bigr)\notag\\
=&0.
\end{align}
The first term vanishes for static spacetimes because the extrinsic curvature of the constant-time hypersurfaces vanishes. In particular, using
\begin{align}
R_{ijkM}h^i{}_{l}h^j{}_{m}h^k{}_{n}u^M=\tilde{\nabla}_lH_{mn}-\tilde{\nabla}_{m}H_{ln},
\end{align}
where $\tilde{\nabla}_i$ denotes the covariant derivative on the constant-time hypersurface, this term vanishes when $H_{\alpha\beta}=0$. The second term vanishes because $u^\mu h_{\mu\nu}=0$ and $u^\mu\nabla_\nu u_\mu=0$. Finally, the third term vanishes by virtue of the Bianchi identity.

Therefore, the action (\ref{action lovelock}) can be rewritten as follows:
\begin{align}
I=&-\frac{1}{16\pi G_n}\int_{M^+\cup M^-}d^{n}x N\mathcal{H}+\int_{M_{n-1}}d^{n-1}x\sqrt{-q}\Biggl(-\sigma+\sum_m\frac{c_m}{16\pi G_N}\Biggl[Q_m\notag\\
&-\sum_m\frac{4m}{2^m}(2m-1)!\ c_mh^{N_1}_{\ \ [L_1}\cdots h^{M_m}_{\ \ K_m}h^{N_m}_{\ \ L_m]}n_{N_1}h^{N L_1}u^M\nabla_M u_N R_{M_2N_2}^{\ \ \ K_2L_2}\cdots R_{M_mN_m}^{\ \ \ K_mL_m}\Biggr]^-\Biggr),
\end{align}
where we have used Stokes' theorem and $n_M u^M=0$. The boundary term can be further recast in the following form.
\begin{align}
&(2m-1)!\ h^{j_1}_{\ \ [l_1}\cdots h^{i_m}_{\ \ k_m}h^{j_m}_{\ \ l_m]}n_{j_1}h^{M l_1}u^N\nabla_N u_M R_{i_2j_2}^{\ \ \ k_2l_2}\cdots R_{i_mj_m}^{\ \ \ k_ml_m}\notag\\
=&\ (2m-2)!\ \gamma^{X_2}_{\ \ [Z_2}\cdots \gamma^{X_m}_{\ \ Z_m}\gamma^{Y_m}_{\ \ W_m]}u^N n^{M}\nabla_N u_M \tilde{R}_{X_2Y_2}^{\ \ \ Z_2W_2}\cdots \tilde{R}_{X_mY_m}^{\ \ \ Z_mW_m}\notag\\
&-2(m-1)(2m-2)!\ \gamma^{X_2}_{\ \ [Z_2}\cdots \gamma^{X_m}_{\ \ Z_m}\gamma^{Y_m}_{\ \ W_m]}u^M\nabla_M u^{W_m}\sum_{k=0}^{m-2}\binom{m-2}{k}(-2)^kK^{Z_2}_{\ \ X_2}K^{W_2}_{\ \ Y_2}\cdots K^{Z_{k+1}}_{\ \ X_{k+1}}K^{W_{k+1}}_{\ \ Y_{k+1}}\notag\\
&\qquad \times{}^{(n-2)}R_{a_{k+2}b_{k+2}}^{\ \ \ c_{k+2}d_{k+2}}\cdots {}^{(n-2)}R_{a_{m-1}b_{m-1}}^{\ \ \ c_{m-1}d_{m-1}}\tilde{R}_{a_mb_m}^{\ \ \ c_mN}n_N\notag\\
=&-K_{MN}u^M u^{N}(2m-2)!\ \gamma^{X_2}_{\ \ [Z_2}\cdots \gamma^{X_m}_{\ \ Z_m}\gamma^{Y_m}_{\ \ W_m]} \sum_{k=0}^{m-1}\binom{m-1}{k}(-2)^kK^{Z_2}_{\ \ X_2}K^{W_2}_{\ \ Y_2}\cdots K^{Z_{k+1}}_{\ \ X_{k+1}}K^{W_{k+1}}_{\ \ Y_{k+1}}{}^{(n-2)}R_{X_{k+2}Y_{k+2}}^{\ \ \ Z_{k+2}W_{k+2}}\notag\\
&\cdots {}^{(n-2)}R_{X_{m}Y_{m}}^{\ \ \ Z_{m}W_{m}}\notag\\
&-2(m-1)(2m-2)!\ \gamma^{X_2}_{\ \ [Z_2}\cdots \gamma^{X_m}_{\ \ Z_m}\gamma^{Y_m}_{\ \ W_m]}u^N\nabla_N u^{W_m}\sum_{k=0}^{m-2}\binom{m-2}{k}(-2)^kK^{Z_2}_{\ \ X_2}K^{W_2}_{\ \ Y_2}\cdots K^{Z_{k+1}}_{\ \ X_{k+1}}K^{W_{k+1}}_{\ \ Y_{k+1}}\notag\\
&\qquad \times{}^{(n-2)}R_{X_{k+2}Y_{k+2}}^{\ \ \ Z_{k+2}W_{k+2}}\cdots {}^{(n-2)}R_{X_{m-1}Y_{m-1}}^{\ \ \ Z_{m-1}W_{m-1}}\tilde{R}_{X_mY_m}^{\ \ \ Z_mM}n_M\notag\\
=&-K_{MN}u^M u^{N}(2m-2)!\ \gamma^{X_2}_{\ \ [Z_2}\cdots \gamma^{X_m}_{\ \ Z_m}\gamma^{Y_m}_{\ \ W_m]} \sum_{k=0}^{m-1}\binom{m-1}{k}(-2)^kK^{Z_2}_{\ \ X_2}K^{W_2}_{\ \ Y_2}\cdots K^{Z_{k+1}}_{\ \ X_{k+1}}K^{W_{k+1}}_{\ \ Y_{k+1}}{}^{(n-2)}R_{X_{k+2}Y_{k+2}}^{\ \ \ Z_{k+2}W_{k+2}}\notag\\
&\cdots {}^{(n-2)}R_{X_{m}Y_{m}}^{\ \ \ Z_{m}W_{m}}\notag\\
&+4(m-1)(2m-2)!\ \gamma^{X_2}_{\ \ [Z_2}\cdots \gamma^{X_m}_{\ \ Z_m}\gamma^{Y_m}_{\ \ W_m]}u^N\nabla_N u^{W_m}\sum_{k=0}^{m-2}\binom{m-2}{k}(-2)^kK^{Z_2}_{\ \ X_2}K^{W_2}_{\ \ Y_2}\cdots K^{Z_{k+1}}_{\ \ X_{k+1}}K^{W_{k+1}}_{\ \ Y_{k+1}}\notag\\
&\qquad \times{}^{(n-2)}R_{X_{k+2}Y_{k+2}}^{\ \ \ Z_{k+2}W_{k+2}}\cdots {}^{(n-2)}R_{X_{m-1}Y_{m-1}}^{\ \ \ Z_{m-1}W_{m-1}}{}^{(n-2)}\nabla_{Y_m}{}^{(d-2)}K^{Z_m}_{\ \ X_m},
\label{relation1}
\end{align}
where ${}^{(n-2)}\nabla_X$ represents the covariant derivative induced on the constant-time hypersurface of the brane $\Sigma_{n-1}$, in the first equality, we decompose the indices into components normal to the brane and components tangent to the brane, and separate the terms according to which indices are associated with the $n_M$ direction. We also use the fact that the extrinsic curvature of the constant-time hypersurfaces vanishes in the present setup and rewrite the resulting geometric quantities in terms of those induced on the constant-time hypersurface of the brane; in the second equality, we similarly rewrite the first term in terms of geometric quantities induced on the constant-time hypersurface of the brane; and in the third equality, we perform the transformation using the following relation.
\begin{align}
\gamma^{X_1}_{\ \ X_2}\gamma^{Y_1}_{\ \ Y_2}\gamma^{Z_1}_{\ \ Z_2}n^M \tilde{R}_{X_1Y_1Z_1M}={}^{(n-2)}\nabla_{X_2}{}^{(n-2)}K_{Y_2Z_2}-{}^{(n-2)}\nabla_{Y_2}{}^{(n-2)}K_{X_2Z_2}.
\end{align}

The second term can be further reduced to the following form.
\begin{align}
&4(m-1)(2m-2)!\ \gamma^{X_2}_{\ \ [Z_2}\cdots \gamma^{X_m}_{\ \ Z_m}\gamma^{Y_m}_{\ \ W_m]}u^M\nabla_M u^{W_m}\sum_{k=0}^{m-2}\binom{m-2}{k}(-2)^kK^{Z_2}_{\ \ X_2}K^{W_2}_{\ \ Y_2}\cdots K^{Z_{k+1}}_{\ \ X_{k+1}}K^{W_{k+1}}_{\ \ Y_{k+1}}\notag\\
&\qquad \times{}^{(n-2)}R_{X_{k+2}Y_{k+2}}^{\ \ \ Z_{k+2}W_{k+2}}\cdots {}^{(n-2)}R_{X_{m-1}Y_{m-1}}^{\ \ \ Z_{m-1}W_{m-1}}{}^{(n-2)}\nabla_{Y_m}{}^{(n-2)}K^{Z_m}_{\ \ X_m}\notag\\
=&\ 4(m-1)(2m-2)!\ \gamma^{X_2}_{\ \ [Z_2}\cdots \gamma^{X_m}_{\ \ Z_m}\gamma^{Y_m}_{\ \ W_m]}u^M\nabla_M u^{W_m}{}^{(n-2)}\nabla_{Y_m}\Bigl(\sum_{k=0}^{m-2}\binom{m-2}{k}\frac{(-2)^k}{2k+1}{}^{(n-2)}K^{Z_2}_{\ \ X_2}{}^{(n-2)}K^{W_2}_{\ \ Y_2}\notag\\
&\qquad \cdots {}^{(n-2)}K^{Z_{k+1}}_{\ \ X_{k+1}}{}^{(n-2)}K^{W_{k+1}}_{\ \ Y_{k+1}}{}^{(n-2)}R_{X_{k+2}Y_{k+2}}^{\ \ \ Z_{k+2}W_{k+2}}\cdots {}^{(n-2)}R_{X_{m-1}Y_{m-1}}^{\ \ \ Z_{m-1}W_{m-1}}{}^{(n-2)}K^{Z_m}_{\ \ X_m}\Bigr)\notag\\
=&-4(m-1)(2m-2)!\ \gamma^{X_2}_{\ \ [Z_2}\cdots \gamma^{X_m}_{\ \ Z_m}\gamma^{Y_m}_{\ \ W_m]}\sum_{k=0}^{m-2}\binom{m-2}{k}\frac{(-2)^k}{2k+1}K^{Z_2}_{\ \ X_2}K^{W_2}_{\ \ Y_2}\cdots K^{Z_{k+1}}_{\ \ X_{k+1}}K^{W_{k+1}}_{\ \ Y_{k+1}}\notag\\
&\qquad \times{}^{(n-2)}R_{X_{k+2}Y_{k+2}}^{\ \ \ Z_{k+2}W_{k+2}}\cdots {}^{(n-2)}R_{X_{m-1}Y_{m-1}}^{\ \ \ Z_{m-1}W_{m-1}}K^{Z_m}_{\ \ X_m}{}^{(n-2)}\gamma^\mu_{\ \ Y_{m}}{}^{(n-2)}\gamma^{\nu W_{m}}D_{\rho}(q^\rho_{\ \ \mu}u^{\sigma}D_\sigma u_\nu-u^\rho D_\mu u_\nu)\notag\\
&+D_{\mu}\Bigl(4(m-1)(2m-2)!\ \gamma^{X_2}_{\ \ [Z_2}\cdots \gamma^{X_m}_{\ \ Z_m}\gamma^{\mu}_{\ \ W_m]}u^M\nabla_M u^{W_m}\sum_{k=0}^{m-2}\binom{m-2}{k}\frac{(-2)^k}{2k+1}K^{Z_2}_{\ \ X_2}K^{W_2}_{\ \ Y_2}\cdots K^{Z_{k+1}}_{\ \ X_{k+1}}K^{W_{k+1}}_{\ \ Y_{k+1}}\notag\\
&\qquad \times{}^{(n-2)}R_{X_{k+2}Y_{k+2}}^{\ \ \ Z_{k+2}W_{k+2}}\cdots {}^{(n-2)}R_{X_{m-1}Y_{m-1}}^{\ \ \ Z_{m-1}W_{m-1}}{}^{(n-2)}K^{Z_m}_{\ \ X_m}\Bigr)\notag\\
=&-2(2m-1)!\ q^{\nu_1}_{\ \ [\sigma_1}q^{\mu_2}_{\ \ \rho_2}\cdots q^{\mu_m}_{\ \ \rho_m}q^{\nu_m}_{\ \ \sigma_m]}\sum_{k=0}^{m-1}\binom{m-1}{k}\frac{(-2)^k}{2k+1}(m-k-1)K^{\rho_2}_{\ \ \mu_2}K^{\sigma_2}_{\ \ \nu_2}\cdots K^{\rho_{k+1}}_{\ \ \mu_{k+1}}K^{\sigma_{k+1}}_{\ \ \nu_{k+1}}\notag\\
&\qquad \times{}^{(n-1)}R_{\mu_{k+2}\nu_{k+2}}^{\ \ \ \rho_{k+2}\sigma_{k+2}}\cdots {}^{(n-1)}R_{\mu_{m-1}\nu_{m-1}}^{\ \ \ \rho_{m-1}\sigma_{m-1}}K^{\rho_m}_{\ \ \mu_m}{}^{(n-1)}R_{\nu_m\nu_1}^{\ \ \ \sigma_m\mu}u_\mu u^{\sigma_1}\notag\\
&+D_{\mu}\Bigl(4(m-1)(2m-2)!\ \gamma^{X_2}_{\ \ [Z_2}\cdots \gamma^{X_m}_{\ \ Z_m}\gamma^{\mu}_{\ \ W_m]}u^\nu\nabla_\nu u^{W_m}\sum_{k=0}^{m-2}\binom{m-2}{k}\frac{(-2)^k}{2k+1}K^{Z_2}_{\ \ X_2}K^{W_2}_{\ \ Y_2}\cdots K^{Z_{k+1}}_{\ \ X_{k+1}}K^{W_{k+1}}_{\ \ Y_{k+1}}\notag\\
&\qquad \times{}^{(n-2)}R_{X_{k+2}Y_{k+2}}^{\ \ \ Z_{k+2}W_{k+2}}\cdots {}^{(n-2)}R_{X_{m-1}Y_{m-1}}^{\ \ \ Z_{m-1}W_{m-1}}{}^{(n-2)}K^{Z_m}_{\ \ X_m}\Bigr),
\end{align}
where $D_\mu$ denotes the covariant derivative on the brane $M_{n-1}$. In the first equality, we use the Bianchi identity and the following relation,
\begin{align}
K^{Z_2}_{\ \ X_2}K^{W_2}_{\ \ Y_2}\cdots K^{Z_{k+1}}_{\ \ X_{k+1}}{}^{(n-2)}\nabla_{Y_m}K^{W_{k+1}}_{\ \ Y_{k+1}}=\frac{1}{2k+1}{}^{(n-2)}\nabla_{Y_m}\left(K^{Z_2}_{\ \ X_2}K^{W_2}_{\ \ Y_2}\cdots K^{Z_{k+1}}_{\ \ X_{k+1}}K^{W_{k+1}}_{\ \ Y_{k+1}}\right).
\label{relation2}
\end{align}
In the second equality, we decompose the indices into components normal and tangent to the brane. We also use the fact that the extrinsic curvature of the constant-time hypersurfaces vanishes in the present setup and rewrite the resulting geometric quantities in terms of those induced on the constant-time hypersurface of the brane. In the third equality, we use the following relation,
\begin{align}
{}^{(n-1)}R_{X\nu Y\mu}u_\mu u^{\nu}={}^{(n-2)}\gamma^\mu_{\ \ X}{}^{(n-2)}\gamma^{\nu}_{\ \ Y}D_{\rho}(q^\rho_{\ \ \mu}u^{\sigma}D_\sigma u_\nu-u^\rho D_\mu u_\nu).
\end{align}

Therefore, the Euclidean action for the spacetime with a cutoff at the lapse function $N=\epsilon$ is given by
\begin{align}
I_{\rm E,\epsilon}=&\beta_0\int_{\Sigma_{n-1,\epsilon}}d^{n-2}xN\sqrt{\gamma}\Bigl(\sigma-\sum_m\frac{c_m}{16\pi G_N}\Biggl[Q_m\notag\\
&-\frac{4m}{2^m}(2m-1)!\ h^{j_1}_{\ \ [l_1}\cdots h^{i_m}_{\ \ k_m}h^{j_m}_{\ \ l_m]}n_{j_1}h^{\mu l_1}u^\nu\nabla_\nu u_\mu R_{i_2j_2}^{\ \ \ k_2l_2}\cdots R_{i_mj_m}^{\ \ \ k_ml_m}\Biggr]^-\Biggr)\notag\\
&+\frac{\beta_0}{16\pi G_n}\int_{\Gamma^+_\epsilon\cup\Gamma^-_\epsilon}d^{n-2}xN\sqrt{G}\sum_m\frac{4m}{2^m}(2m-1)!\ c_mh^{j_1}_{\ \ [l_1}\cdots h^{i_m}_{\ \ k_m}h^{j_m}_{\ \ l_m]}r_{j_1}h^{\mu l_1}u^\nu\nabla_\nu u_\mu R_{i_2j_2}^{\ \ \ k_2l_2}\cdots R_{i_mj_m}^{\ \ \ k_ml_m}\notag\\
=&\beta_0\int_{\Sigma_{n-1,\epsilon}}d^{n-2}xN\sqrt{\gamma}\Bigl(\sigma-\sum_m\frac{c_m}{16\pi G_N}\Biggl[Q_m\notag\\
&+\frac{4m}{2^m}\gamma^{X_1Y_1X_2Y_2\cdots X_{m-1}Y_{m-1}}_{Z_1W_1Z_2W_2\cdots Z_{m-1}W_{m-1}}\sum_{k=0}^{m-1}\binom{m-1}{k}(-2)^kK_{\mu\nu}u^\mu u^\nu K^{Z_1}_{\ \ X_1}K^{W_1}_{\ \ Y_1}\cdots K^{Z_k}_{\ \ X_k}K^{W_k}_{\ \ Y_k}\notag\\
&\times{}^{(n-1)}R_{X_{k+1}Y_{k+1}}^{\ \ \ Z_{k+1}W_{k+1}}\cdots{}^{(n-1)}R_{X_{m-1}Y_{m-1}}^{\ \ \ Z_{m-1}W_{m-1}}\notag\\
&+\frac{8m}{2^m}q^{\mu_1\nu_1\mu_2\nu_2\cdots \mu_{m-1}\nu_{m-1}\mu_m}_{\rho_1\sigma_1\rho_2\sigma_2\cdots \rho_{m-1}\sigma_{m-1}c_m}\sum_{k=0}^{m-1}\binom{m-1}{k}\frac{(-2)^k}{2k+1}(m-k-1)K^{\rho_1}_{\ \ \mu_1}K^{\sigma_1}_{\ \ \nu_1}\cdots K^{\rho_{k}}_{\ \ \mu_{k}}K^{\sigma_{k}}_{\ \ \nu_{k}}\notag\\
&\qquad \times{}^{(n-1)}R_{\mu_{k+1}\nu_{k+1}}^{\ \ \ \rho_{k+1}\sigma_{k+1}}\cdots {}^{(n-1)}R_{\mu_{m-2}\nu_{m-2}}^{\ \ \ \rho_{m-2}\sigma_{m-2}}K^{\rho_m}_{\ \ \mu_m}{}^{(n-1)}R_{\mu_{m-1}\nu_{m-1}}^{\ \ \ \rho_{m-1}\mu}u_\mu u^{\sigma_{m-1}}\Biggr]^-\Bigr)\notag\\
&+\frac{\beta_0}{16\pi G_n}\int_{\Gamma^+_\epsilon\cup\Gamma^-_\epsilon}d^{n-2}xN\sqrt{G}\sum_m\frac{4m}{2^m}(2m-1)!\ c_mh^{j_1}_{\ \ [l_1}\cdots h^{i_m}_{\ \ k_m}h^{j_m}_{\ \ l_m]}r_{j_1}h^{\mu l_1}u^\nu\nabla_\nu u_\mu R_{i_2j_2}^{\ \ \ k_2l_2}\cdots R_{i_mj_m}^{\ \ \ k_ml_m}\notag\\
&+\frac{\beta_0}{16\pi G_n}\int_{\partial\Gamma_\epsilon}dx^{n-3}xN\sqrt{p}\Biggl[\sum_m\frac{16m(m-1)}{2^m}c_m\gamma^{X_1Y_1X_2Y_2\cdots X_{m-1}Y_{m-1}}_{Z_1W_1Z_2W_2\cdots Z_{m-1}W_{m-1}}u^\mu\nabla_\mu u^{W_{m-1}}r_{Y_{m-1}}\sum_{k=0}^{m-2}\binom{m-2}{k}\frac{(-2)^k}{2k+1}K^{Z_1}_{\ \ X_1}K^{W_1}_{\ \ Y_1}\notag\\
&\cdots K^{Z_{k}}_{\ \ X_{k}}K^{W_{k}}_{\ \ Y_{k}}{}^{(n-2)}R_{X_{k+1}Y_{k+1}}^{\ \ \ Z_{k+1}W_{k+1}}\cdots {}^{(n-2)}R_{X_{m-2}Y_{m-2}}^{\ \ \ Z_{m-2}W_{m-2}}{}^{(n-2)}K^{Z_{m-1}}_{\ \ X_{m-1}}\Biggr]^-\notag\\
=&\ \frac{\beta_0}{16\pi G_n}\int_{\Gamma^+_\epsilon\cup\Gamma^-_\epsilon}d^{n-2}xN\sqrt{G}\sum_m\frac{4m}{2^m}(2m-1)!\ c_mh^{j_1}_{\ \ [l_1}\cdots h^{i_m}_{\ \ k_m}h^{j_m}_{\ \ l_m]}r_{j_1}h^{\mu l_1}u^\nu\nabla_\nu u_\mu R_{i_2j_2}^{\ \ \ k_2l_2}\cdots R_{i_mj_m}^{\ \ \ k_ml_m}\notag\\
&+\frac{\beta_0}{16\pi G_n}\int_{\partial\Gamma_\epsilon}dx^{n-3}xN\sqrt{p}\Biggl[\sum_m\frac{16m(m-1)}{2^m}c_m\gamma^{X_1Y_1X_2Y_2\cdots X_{m-1}Y_{m-1}}_{Z_1W_1Z_2W_2\cdots Z_{m-1}W_{m-1}}u^\mu\nabla_\mu u^{W_{m-1}}r_{Y_{m-1}}\sum_{k=0}^{m-2}\binom{m-2}{k}\frac{(-2)^k}{2k+1}K^{Z_1}_{\ \ X_1}K^{W_1}_{\ \ Y_1}\notag\\
&\cdots K^{Z_{k}}_{\ \ X_{k}}K^{W_{k}}_{\ \ Y_{k}}{}^{(n-2)}R_{X_{k+1}Y_{k+1}}^{\ \ \ Z_{k+1}W_{k+1}}\cdots {}^{(n-2)}R_{X_{m-2}Y_{m-2}}^{\ \ \ Z_{m-2}W_{m-2}}{}^{(n-2)}K^{Z_{m-1}}_{\ \ X_{m-1}}\Biggr]^-,
\end{align}
where, in the second equality, we have used Eqs.~(\ref{relation1}) and (\ref{relation2}), while in the third equality, the terms on $\Sigma_{n-1,\epsilon}$ vanish as a consequence of the junction conditions, and the tensor $\gamma^{X_1Y_1X_2Y_2\cdots X_{m-1}Y_{m-1}}*{Z_1W_1Z_2W_2\cdots Z*{m-1}W_{m-1}}$ is defined as
\begin{align}
\gamma^{X_1Y_1X_2Y_2\cdots X_{m-1}Y_{m-1}}_{Z_1W_1Z_2W_2\cdots Z_{m-1}W_{m-1}}:=(2m-2)!\ \gamma^{X_1}_{\ \ [Z_1}\cdots \gamma^{X_{m-1}}_{\ \ Z_{m-1}}\gamma^{Y_{m-1}}_{\ \ W_{m-1}]}.
\end{align}

Then, the dSitter entropy is
\begin{align}
S_{\rm dS}=&-\lim_{\epsilon\rightarrow0}I_{\rm E,\epsilon}\notag\\
=&\lim_{\epsilon\rightarrow0}\Biggl(\frac{1}{4G_n}\int_{\Gamma^+_\epsilon\cup\Gamma^-_\epsilon}d^{n-2}x\sqrt{G}\sum_m\frac{m}{2^{m-1}}(2m-1)!\ c_mh^{j_1}_{\ \ [l_1}\cdots h^{i_m}_{\ \ k_m}h^{j_m}_{\ \ l_m]}r_{j_1}r^{l_1} R_{i_2j_2}^{\ \ \ k_2l_2}\cdots R_{i_mj_m}^{\ \ \ k_ml_m}\notag\\
&+\frac{1}{4G_n}\int_{\partial\Gamma_\epsilon}dx^{n-3}x\sqrt{p}\Biggl[\sum_m\frac{4m(m-1)}{2^{m-1}}c_m\gamma^{X_1Y_1X_2Y_2\cdots X_{m-1}Y_{m-1}}_{Z_1W_1Z_2W_2\cdots Z_{m-1}W_{m-1}}r^{W_{m-1}}r_{Y_{m-1}}\sum_{k=0}^{m-2}\binom{m-2}{k}\frac{(-2)^k}{2k+1}K^{Z_1}_{\ \ X_1}K^{W_1}_{\ \ Y_1}\notag\\
&\cdots K^{Z_{k}}_{\ \ X_{k}}K^{W_{k}}_{\ \ Y_{k}}{}^{(n-2)}R_{X_{k+1}Y_{k+1}}^{\ \ \ Z_{k+1}W_{k+1}}\cdots {}^{(n-2)}R_{X_{m-2}Y_{m-2}}^{\ \ \ Z_{m-2}W_{m-2}}{}^{(n-2)}K^{Z_{m-1}}_{\ \ X_{m-1}}\Biggr]^-\Biggr)\notag\\
=&\ \frac{1}{4G_n}\int_{\Gamma^+\cup\Gamma^-}d^{n-2}x\sqrt{G}\sum_m\frac{m}{2^{m-1}}c_mG^{x_1y_1\cdots x_{m-1}y_{m-1}}_{z_1w_1\cdots z_{m-1}w_{m-1}}{}^{(n-2)}\tilde{R}_{a_2b_2}^{\ \ \ c_2d_2}\cdots {}^{(n-2)}\tilde{R}_{a_mb_m}^{\ \ \ c_md_m}\notag\\
&+\frac{1}{4G_n}\int_{\partial\Gamma}dx^{n-3}xN\sqrt{p}\Biggl[\sum_m\frac{4m(m-1)}{2^{m-1}}c_mp^{A_1B_1\cdots A_{m-1}}_{C_1D_1\cdots C_{m-1}}\sum_{k=0}^{m-2}\binom{m-2}{k}\frac{(-2)^k}{2k+1}{}^{(n-3)}K^{C_1}_{\ \ A_1}{}^{(n-3)}K^{D_1}_{\ \ B_1}\notag\\
&\cdots {}^{(n-3)}K^{C_{k}}_{\ \ A_{k}}{}^{(n-3)}K^{D_{k}}_{\ \ B_{k}}{}^{(n-3)}R_{A_{k+1}B_{k+1}}^{\ \ \ C_{k+1}D_{k+1}}\cdots {}^{(n-3)}R_{A_{m-2}B_{m-2}}^{\ \ \ C_{m-2}D_{m-2}}{}^{(n-3)}K^{C_{m-1}}_{\ \ A_{m-1}}\Biggr]^-\notag\\
=&\ S_{\rm JM},
\end{align}
where, in the third equality, we have used the decomposition of both the first and second terms into components along the direction of the unit normal vector $r_{\mu}$ to the horizon and those orthogonal to it, together with the fact that the extrinsic curvature associated with $r_{\mu}$ vanishes on the horizon.

Therefore, we have shown that the de Sitter entropy coincides with the holographic entanglement entropy for static spacetimes even in general Lovelock gravity.

\section{Conclusion and Discussion}
\label{sec:conclusion}

In this work, we have investigated the relation between de Sitter entropy and holographic entanglement entropy in Gauss-Bonnet gravity. We first formulated Gauss-Bonnet gravity in the canonical formalism and derived the de Sitter entropy from the Euclidean gravitational action, including the contributions from the Gauss-Bonnet surface terms. We then compared this result with the holographic entanglement entropy obtained from the Jacobson-Myers functional for general stationary asymptotically de Sitter braneworld spacetimes in the RS II model.

Our analysis shows that the de Sitter entropy and the holographic entanglement entropy do not coincide in general for stationary spacetimes. The discrepancy originates from an extrinsic-curvature contribution in the entropy expressions. We have therefore identified the geometrical contribution responsible for the inequivalence between the two entropy prescriptions. In contrast, for static spacetimes, the relevant extrinsic-curvature contribution vanishes, and the two entropy prescriptions reduce to the same expression. This also shows that the agreement in static asymptotically de Sitter braneworld spacetimes is directly tied to the vanishing of the extrinsic-curvature contribution. Thus, the equality between de Sitter entropy and holographic entanglement entropy holds generally for static spacetimes, but does not extend to general stationary spacetimes.

We have further extended our analysis to general Lovelock gravity and shown that the de Sitter entropy coincides with the holographic entanglement entropy for static spacetimes also in the general Lovelock case. This demonstrates that the agreement found in Gauss--Bonnet gravity is not restricted to the quadratic Lovelock term, but persists for the full class of Lovelock theories for static spacetimes. This agreement arises because the extrinsic curvature of constant-time hypersurfaces vanishes for static spacetimes, and therefore the de Sitter entropy and the holographic entanglement entropy coincide also in general Lovelock gravity.

The same conclusion applies to braneworld black holes. Previous studies have investigated the relation between the black hole entropy and holographic entanglement entropy in braneworld black holes \cite{Emparan:2006ni}, providing a basis for considering the corresponding entropy relation in the present framework. We have also investigated the relation between the Gibbons-Hawking entropy and the holographic entanglement entropy for braneworld black holes. In this case, the corresponding gravitational entropy is the Gibbons-Hawking entropy associated with the event horizon. The calculation relating the Gibbons-Hawking entropy to the holographic entanglement entropy is essentially identical to that for the de Sitter entropy considered above. The only distinction is the horizon on which the entropy is evaluated: the de Sitter entropy is evaluated on the cosmological horizon, whereas the Gibbons-Hawking entropy of a black hole is evaluated on the event horizon. Since the geometrical structure of the entropy calculation is otherwise the same, the same result follows for braneworld black holes: the Gibbons-Hawking entropy and the holographic entanglement entropy generally differ for stationary spacetimes, while they coincide for static spacetimes. This demonstrates that the mismatch identified here is not specific to the cosmological horizon, but also arises for black hole horizons in braneworld gravity.

Although the geometrical origin of the discrepancy can be identified, its physical interpretation remains an open question. The agreement between de Sitter entropy and holographic entanglement entropy for static spacetimes suggests a close relation between these two notions of gravitational entropy. It is therefore intriguing that this agreement is lost for general stationary spacetimes. In particular, it remains unclear why the presence of a nonvanishing extrinsic-curvature contribution should lead to physically distinct entropy prescriptions, and what fundamental distinction between de Sitter entropy and holographic entanglement entropy is reflected by this difference.

Our results provide a concrete setting in which the physical relation between gravitational entropy and holographic entanglement entropy can be further investigated. An important direction for future work is to clarify the physical origin of the extrinsic-curvature contribution and to determine whether the mismatch found here is a generic feature of stationary spacetimes in higher-curvature gravity. Extending the analysis to other stationary geometries and more general higher-curvature theories may provide further insight into the relation between gravitational entropy, extrinsic curvature, and holographic entanglement entropy.

\section*{Acknowledgments}

The author would like to thank Keisuke Izumi, Tetsuya Shiromizu and Daisuke Yoshida for useful discussions and valuable comments.

\bibliography{references}

\begin{thebibliography}{37}%
\makeatletter
\providecommand \@ifxundefined [1]{%
 \@ifx{#1\undefined}
}%
\providecommand \@ifnum [1]{%
 \ifnum #1\expandafter \@firstoftwo
 \else \expandafter \@secondoftwo
 \fi
}%
\providecommand \@ifx [1]{%
 \ifx #1\expandafter \@firstoftwo
 \else \expandafter \@secondoftwo
 \fi
}%
\providecommand \natexlab [1]{#1}%
\providecommand \enquote  [1]{``#1''}%
\providecommand \bibnamefont  [1]{#1}%
\providecommand \bibfnamefont [1]{#1}%
\providecommand \citenamefont [1]{#1}%
\providecommand \href@noop [0]{\@secondoftwo}%
\providecommand \href [0]{\begingroup \@sanitize@url \@href}%
\providecommand \@href[1]{\@@startlink{#1}\@@href}%
\providecommand \@@href[1]{\endgroup#1\@@endlink}%
\providecommand \@sanitize@url [0]{\catcode `\\12\catcode `\$12\catcode
  `\&12\catcode `\#12\catcode `\^12\catcode `\_12\catcode `\%12\relax}%
\providecommand \@@startlink[1]{}%
\providecommand \@@endlink[0]{}%
\providecommand \url  [0]{\begingroup\@sanitize@url \@url }%
\providecommand \@url [1]{\endgroup\@href {#1}{\urlprefix }}%
\providecommand \urlprefix  [0]{URL }%
\providecommand \Eprint [0]{\href }%
\providecommand \doibase [0]{https://doi.org/}%
\providecommand \selectlanguage [0]{\@gobble}%
\providecommand \bibinfo  [0]{\@secondoftwo}%
\providecommand \bibfield  [0]{\@secondoftwo}%
\providecommand \translation [1]{[#1]}%
\providecommand \BibitemOpen [0]{}%
\providecommand \bibitemStop [0]{}%
\providecommand \bibitemNoStop [0]{.\EOS\space}%
\providecommand \EOS [0]{\spacefactor3000\relax}%
\providecommand \BibitemShut  [1]{\csname bibitem#1\endcsname}%
\let\auto@bib@innerbib\@empty
\bibitem [{\citenamefont {Hawking}(1974)}]{Hawking:1974rv}%
  \BibitemOpen
  \bibfield  {author} {\bibinfo {author} {\bibfnamefont {S.~W.}\ \bibnamefont
  {Hawking}},\ }\bibfield  {title} {\bibinfo {title} {{Black hole
  explosions}},\ }\href {https://doi.org/10.1038/248030a0} {\bibfield
  {journal} {\bibinfo  {journal} {Nature}\ }\textbf {\bibinfo {volume} {248}},\
  \bibinfo {pages} {30} (\bibinfo {year} {1974})}\BibitemShut {NoStop}%
\bibitem [{\citenamefont {Hawking}(1975)}]{Hawking:1975vcx}%
  \BibitemOpen
  \bibfield  {author} {\bibinfo {author} {\bibfnamefont {S.~W.}\ \bibnamefont
  {Hawking}},\ }\bibfield  {title} {\bibinfo {title} {{Particle Creation by
  Black Holes}},\ }\href {https://doi.org/10.1007/BF02345020} {\bibfield
  {journal} {\bibinfo  {journal} {Commun. Math. Phys.}\ }\textbf {\bibinfo
  {volume} {43}},\ \bibinfo {pages} {199} (\bibinfo {year} {1975})},\ \bibinfo
  {note} {[Erratum: Commun.Math.Phys. 46, 206 (1976)]}\BibitemShut {NoStop}%
\bibitem [{\citenamefont {Bekenstein}(1972)}]{Bekenstein:1972tm}%
  \BibitemOpen
  \bibfield  {author} {\bibinfo {author} {\bibfnamefont {J.~D.}\ \bibnamefont
  {Bekenstein}},\ }\bibfield  {title} {\bibinfo {title} {{Black holes and the
  second law}},\ }\href {https://doi.org/10.1007/BF02757029} {\bibfield
  {journal} {\bibinfo  {journal} {Lett. Nuovo Cim.}\ }\textbf {\bibinfo
  {volume} {4}},\ \bibinfo {pages} {737} (\bibinfo {year} {1972})}\BibitemShut
  {NoStop}%
\bibitem [{\citenamefont {Bekenstein}(1973)}]{Bekenstein:1973ur}%
  \BibitemOpen
  \bibfield  {author} {\bibinfo {author} {\bibfnamefont {J.~D.}\ \bibnamefont
  {Bekenstein}},\ }\bibfield  {title} {\bibinfo {title} {{Black holes and
  entropy}},\ }\href {https://doi.org/10.1103/PhysRevD.7.2333} {\bibfield
  {journal} {\bibinfo  {journal} {Phys. Rev. D}\ }\textbf {\bibinfo {volume}
  {7}},\ \bibinfo {pages} {2333} (\bibinfo {year} {1973})}\BibitemShut
  {NoStop}%
\bibitem [{\citenamefont {Bekenstein}(1974)}]{Bekenstein:1974ax}%
  \BibitemOpen
  \bibfield  {author} {\bibinfo {author} {\bibfnamefont {J.~D.}\ \bibnamefont
  {Bekenstein}},\ }\bibfield  {title} {\bibinfo {title} {{Generalized second
  law of thermodynamics in black hole physics}},\ }\href
  {https://doi.org/10.1103/PhysRevD.9.3292} {\bibfield  {journal} {\bibinfo
  {journal} {Phys. Rev. D}\ }\textbf {\bibinfo {volume} {9}},\ \bibinfo {pages}
  {3292} (\bibinfo {year} {1974})}\BibitemShut {NoStop}%
\bibitem [{\citenamefont {Wald}(1993)}]{Wald:1993nt}%
  \BibitemOpen
  \bibfield  {author} {\bibinfo {author} {\bibfnamefont {R.~M.}\ \bibnamefont
  {Wald}},\ }\bibfield  {title} {\bibinfo {title} {{Black hole entropy is the
  Noether charge}},\ }\href {https://doi.org/10.1103/PhysRevD.48.R3427}
  {\bibfield  {journal} {\bibinfo  {journal} {Phys. Rev. D}\ }\textbf {\bibinfo
  {volume} {48}},\ \bibinfo {pages} {R3427} (\bibinfo {year} {1993})},\ \Eprint
  {https://arxiv.org/abs/gr-qc/9307038} {arXiv:gr-qc/9307038} \BibitemShut
  {NoStop}%
\bibitem [{\citenamefont {Jacobson}\ \emph {et~al.}(1994)\citenamefont
  {Jacobson}, \citenamefont {Kang},\ and\ \citenamefont
  {Myers}}]{Jacobson:1993vj}%
  \BibitemOpen
  \bibfield  {author} {\bibinfo {author} {\bibfnamefont {T.}~\bibnamefont
  {Jacobson}}, \bibinfo {author} {\bibfnamefont {G.}~\bibnamefont {Kang}},\
  and\ \bibinfo {author} {\bibfnamefont {R.~C.}\ \bibnamefont {Myers}},\
  }\bibfield  {title} {\bibinfo {title} {{On black hole entropy}},\ }\href
  {https://doi.org/10.1103/PhysRevD.49.6587} {\bibfield  {journal} {\bibinfo
  {journal} {Phys. Rev. D}\ }\textbf {\bibinfo {volume} {49}},\ \bibinfo
  {pages} {6587} (\bibinfo {year} {1994})},\ \Eprint
  {https://arxiv.org/abs/gr-qc/9312023} {arXiv:gr-qc/9312023} \BibitemShut
  {NoStop}%
\bibitem [{\citenamefont {Jacobson}\ and\ \citenamefont
  {Myers}(1993)}]{Jacobson:1993xs}%
  \BibitemOpen
  \bibfield  {author} {\bibinfo {author} {\bibfnamefont {T.}~\bibnamefont
  {Jacobson}}\ and\ \bibinfo {author} {\bibfnamefont {R.~C.}\ \bibnamefont
  {Myers}},\ }\bibfield  {title} {\bibinfo {title} {{Black hole entropy and
  higher curvature interactions}},\ }\href
  {https://doi.org/10.1103/PhysRevLett.70.3684} {\bibfield  {journal} {\bibinfo
   {journal} {Phys. Rev. Lett.}\ }\textbf {\bibinfo {volume} {70}},\ \bibinfo
  {pages} {3684} (\bibinfo {year} {1993})},\ \Eprint
  {https://arxiv.org/abs/hep-th/9305016} {arXiv:hep-th/9305016} \BibitemShut
  {NoStop}%
\bibitem [{\citenamefont {Maldacena}(1998)}]{Maldacena:1997re}%
  \BibitemOpen
  \bibfield  {author} {\bibinfo {author} {\bibfnamefont {J.~M.}\ \bibnamefont
  {Maldacena}},\ }\bibfield  {title} {\bibinfo {title} {{The Large $N$ limit of
  superconformal field theories and supergravity}},\ }\href
  {https://doi.org/10.4310/ATMP.1998.v2.n2.a1} {\bibfield  {journal} {\bibinfo
  {journal} {Adv. Theor. Math. Phys.}\ }\textbf {\bibinfo {volume} {2}},\
  \bibinfo {pages} {231} (\bibinfo {year} {1998})},\ \Eprint
  {https://arxiv.org/abs/hep-th/9711200} {arXiv:hep-th/9711200} \BibitemShut
  {NoStop}%
\bibitem [{\citenamefont {Gubser}\ \emph {et~al.}(1998)\citenamefont {Gubser},
  \citenamefont {Klebanov},\ and\ \citenamefont {Polyakov}}]{Gubser:1998bc}%
  \BibitemOpen
  \bibfield  {author} {\bibinfo {author} {\bibfnamefont {S.~S.}\ \bibnamefont
  {Gubser}}, \bibinfo {author} {\bibfnamefont {I.~R.}\ \bibnamefont
  {Klebanov}},\ and\ \bibinfo {author} {\bibfnamefont {A.~M.}\ \bibnamefont
  {Polyakov}},\ }\bibfield  {title} {\bibinfo {title} {{Gauge theory
  correlators from noncritical string theory}},\ }\href
  {https://doi.org/10.1016/S0370-2693(98)00377-3} {\bibfield  {journal}
  {\bibinfo  {journal} {Phys. Lett. B}\ }\textbf {\bibinfo {volume} {428}},\
  \bibinfo {pages} {105} (\bibinfo {year} {1998})},\ \Eprint
  {https://arxiv.org/abs/hep-th/9802109} {arXiv:hep-th/9802109} \BibitemShut
  {NoStop}%
\bibitem [{\citenamefont {Witten}(1998)}]{Witten:1998qj}%
  \BibitemOpen
  \bibfield  {author} {\bibinfo {author} {\bibfnamefont {E.}~\bibnamefont
  {Witten}},\ }\bibfield  {title} {\bibinfo {title} {{Anti de Sitter space and
  holography}},\ }\href {https://doi.org/10.4310/ATMP.1998.v2.n2.a2} {\bibfield
   {journal} {\bibinfo  {journal} {Adv. Theor. Math. Phys.}\ }\textbf {\bibinfo
  {volume} {2}},\ \bibinfo {pages} {253} (\bibinfo {year} {1998})},\ \Eprint
  {https://arxiv.org/abs/hep-th/9802150} {arXiv:hep-th/9802150} \BibitemShut
  {NoStop}%
\bibitem [{\citenamefont {Ryu}\ and\ \citenamefont
  {Takayanagi}(2006)}]{Ryu:2006bv}%
  \BibitemOpen
  \bibfield  {author} {\bibinfo {author} {\bibfnamefont {S.}~\bibnamefont
  {Ryu}}\ and\ \bibinfo {author} {\bibfnamefont {T.}~\bibnamefont
  {Takayanagi}},\ }\bibfield  {title} {\bibinfo {title} {{Holographic
  derivation of entanglement entropy from AdS/CFT}},\ }\href
  {https://doi.org/10.1103/PhysRevLett.96.181602} {\bibfield  {journal}
  {\bibinfo  {journal} {Phys. Rev. Lett.}\ }\textbf {\bibinfo {volume} {96}},\
  \bibinfo {pages} {181602} (\bibinfo {year} {2006})},\ \Eprint
  {https://arxiv.org/abs/hep-th/0603001} {arXiv:hep-th/0603001} \BibitemShut
  {NoStop}%
\bibitem [{\citenamefont {Hubeny}\ \emph {et~al.}(2007)\citenamefont {Hubeny},
  \citenamefont {Rangamani},\ and\ \citenamefont {Takayanagi}}]{Hubeny:2007xt}%
  \BibitemOpen
  \bibfield  {author} {\bibinfo {author} {\bibfnamefont {V.~E.}\ \bibnamefont
  {Hubeny}}, \bibinfo {author} {\bibfnamefont {M.}~\bibnamefont {Rangamani}},\
  and\ \bibinfo {author} {\bibfnamefont {T.}~\bibnamefont {Takayanagi}},\
  }\bibfield  {title} {\bibinfo {title} {{A Covariant holographic entanglement
  entropy proposal}},\ }\href {https://doi.org/10.1088/1126-6708/2007/07/062}
  {\bibfield  {journal} {\bibinfo  {journal} {JHEP}\ }\textbf {\bibinfo
  {volume} {07}},\ \bibinfo {pages} {062}},\ \Eprint
  {https://arxiv.org/abs/0705.0016} {arXiv:0705.0016 [hep-th]} \BibitemShut
  {NoStop}%
\bibitem [{\citenamefont {Gubser}(2001)}]{Gubser:1999vj}%
  \BibitemOpen
  \bibfield  {author} {\bibinfo {author} {\bibfnamefont {S.~S.}\ \bibnamefont
  {Gubser}},\ }\bibfield  {title} {\bibinfo {title} {{AdS / CFT and gravity}},\
  }\href {https://doi.org/10.1103/PhysRevD.63.084017} {\bibfield  {journal}
  {\bibinfo  {journal} {Phys. Rev. D}\ }\textbf {\bibinfo {volume} {63}},\
  \bibinfo {pages} {084017} (\bibinfo {year} {2001})},\ \Eprint
  {https://arxiv.org/abs/hep-th/9912001} {arXiv:hep-th/9912001} \BibitemShut
  {NoStop}%
\bibitem [{\citenamefont {Giddings}\ \emph {et~al.}(2000)\citenamefont
  {Giddings}, \citenamefont {Katz},\ and\ \citenamefont
  {Randall}}]{Giddings:2000mu}%
  \BibitemOpen
  \bibfield  {author} {\bibinfo {author} {\bibfnamefont {S.~B.}\ \bibnamefont
  {Giddings}}, \bibinfo {author} {\bibfnamefont {E.}~\bibnamefont {Katz}},\
  and\ \bibinfo {author} {\bibfnamefont {L.}~\bibnamefont {Randall}},\
  }\bibfield  {title} {\bibinfo {title} {{Linearized gravity in brane
  backgrounds}},\ }\href {https://doi.org/10.1088/1126-6708/2000/03/023}
  {\bibfield  {journal} {\bibinfo  {journal} {JHEP}\ }\textbf {\bibinfo
  {volume} {03}},\ \bibinfo {pages} {023}},\ \Eprint
  {https://arxiv.org/abs/hep-th/0002091} {arXiv:hep-th/0002091} \BibitemShut
  {NoStop}%
\bibitem [{\citenamefont {Shiromizu}\ and\ \citenamefont
  {Ida}(2001)}]{Shiromizu:2001jm}%
  \BibitemOpen
  \bibfield  {author} {\bibinfo {author} {\bibfnamefont {T.}~\bibnamefont
  {Shiromizu}}\ and\ \bibinfo {author} {\bibfnamefont {D.}~\bibnamefont
  {Ida}},\ }\bibfield  {title} {\bibinfo {title} {{Anti-de Sitter no hair, AdS
  / CFT and the brane world}},\ }\href
  {https://doi.org/10.1103/PhysRevD.64.044015} {\bibfield  {journal} {\bibinfo
  {journal} {Phys. Rev. D}\ }\textbf {\bibinfo {volume} {64}},\ \bibinfo
  {pages} {044015} (\bibinfo {year} {2001})},\ \Eprint
  {https://arxiv.org/abs/hep-th/0102035} {arXiv:hep-th/0102035} \BibitemShut
  {NoStop}%
\bibitem [{\citenamefont {Shiromizu}\ \emph {et~al.}(2002)\citenamefont
  {Shiromizu}, \citenamefont {Torii},\ and\ \citenamefont
  {Ida}}]{Shiromizu:2001ve}%
  \BibitemOpen
  \bibfield  {author} {\bibinfo {author} {\bibfnamefont {T.}~\bibnamefont
  {Shiromizu}}, \bibinfo {author} {\bibfnamefont {T.}~\bibnamefont {Torii}},\
  and\ \bibinfo {author} {\bibfnamefont {D.}~\bibnamefont {Ida}},\ }\bibfield
  {title} {\bibinfo {title} {{Brane world and holography}},\ }\href
  {https://doi.org/10.1088/1126-6708/2002/03/007} {\bibfield  {journal}
  {\bibinfo  {journal} {JHEP}\ }\textbf {\bibinfo {volume} {03}},\ \bibinfo
  {pages} {007}},\ \Eprint {https://arxiv.org/abs/hep-th/0105256}
  {arXiv:hep-th/0105256} \BibitemShut {NoStop}%
\bibitem [{\citenamefont {Nojiri}\ \emph {et~al.}(2000)\citenamefont {Nojiri},
  \citenamefont {Odintsov},\ and\ \citenamefont {Zerbini}}]{Nojiri:2000eb}%
  \BibitemOpen
  \bibfield  {author} {\bibinfo {author} {\bibfnamefont {S.}~\bibnamefont
  {Nojiri}}, \bibinfo {author} {\bibfnamefont {S.~D.}\ \bibnamefont
  {Odintsov}},\ and\ \bibinfo {author} {\bibfnamefont {S.}~\bibnamefont
  {Zerbini}},\ }\bibfield  {title} {\bibinfo {title} {{Quantum (in)stability of
  dilatonic AdS backgrounds and holographic renormalization group with
  gravity}},\ }\href {https://doi.org/10.1103/PhysRevD.62.064006} {\bibfield
  {journal} {\bibinfo  {journal} {Phys. Rev. D}\ }\textbf {\bibinfo {volume}
  {62}},\ \bibinfo {pages} {064006} (\bibinfo {year} {2000})},\ \Eprint
  {https://arxiv.org/abs/hep-th/0001192} {arXiv:hep-th/0001192} \BibitemShut
  {NoStop}%
\bibitem [{\citenamefont {Nojiri}\ and\ \citenamefont
  {Odintsov}(2000)}]{Nojiri:2000gb}%
  \BibitemOpen
  \bibfield  {author} {\bibinfo {author} {\bibfnamefont {S.}~\bibnamefont
  {Nojiri}}\ and\ \bibinfo {author} {\bibfnamefont {S.~D.}\ \bibnamefont
  {Odintsov}},\ }\bibfield  {title} {\bibinfo {title} {{Brane world inflation
  induced by quantum effects}},\ }\href
  {https://doi.org/10.1016/S0370-2693(00)00629-8} {\bibfield  {journal}
  {\bibinfo  {journal} {Phys. Lett. B}\ }\textbf {\bibinfo {volume} {484}},\
  \bibinfo {pages} {119} (\bibinfo {year} {2000})},\ \Eprint
  {https://arxiv.org/abs/hep-th/0004097} {arXiv:hep-th/0004097} \BibitemShut
  {NoStop}%
\bibitem [{\citenamefont {Hawking}\ \emph {et~al.}(2000)\citenamefont
  {Hawking}, \citenamefont {Hertog},\ and\ \citenamefont
  {Reall}}]{Hawking:2000kj}%
  \BibitemOpen
  \bibfield  {author} {\bibinfo {author} {\bibfnamefont {S.~W.}\ \bibnamefont
  {Hawking}}, \bibinfo {author} {\bibfnamefont {T.}~\bibnamefont {Hertog}},\
  and\ \bibinfo {author} {\bibfnamefont {H.~S.}\ \bibnamefont {Reall}},\
  }\bibfield  {title} {\bibinfo {title} {{Brane new world}},\ }\href
  {https://doi.org/10.1103/PhysRevD.62.043501} {\bibfield  {journal} {\bibinfo
  {journal} {Phys. Rev. D}\ }\textbf {\bibinfo {volume} {62}},\ \bibinfo
  {pages} {043501} (\bibinfo {year} {2000})},\ \Eprint
  {https://arxiv.org/abs/hep-th/0003052} {arXiv:hep-th/0003052} \BibitemShut
  {NoStop}%
\bibitem [{\citenamefont {Koyama}\ and\ \citenamefont
  {Soda}(2001)}]{Koyama:2001rf}%
  \BibitemOpen
  \bibfield  {author} {\bibinfo {author} {\bibfnamefont {K.}~\bibnamefont
  {Koyama}}\ and\ \bibinfo {author} {\bibfnamefont {J.}~\bibnamefont {Soda}},\
  }\bibfield  {title} {\bibinfo {title} {{Strongly coupled CFT in FRW universe
  from AdS / CFT correspondence}},\ }\href
  {https://doi.org/10.1088/1126-6708/2001/05/027} {\bibfield  {journal}
  {\bibinfo  {journal} {JHEP}\ }\textbf {\bibinfo {volume} {05}},\ \bibinfo
  {pages} {027}},\ \Eprint {https://arxiv.org/abs/hep-th/0101164}
  {arXiv:hep-th/0101164} \BibitemShut {NoStop}%
\bibitem [{\citenamefont {Kanno}\ and\ \citenamefont
  {Soda}(2002)}]{Kanno:2002iaa}%
  \BibitemOpen
  \bibfield  {author} {\bibinfo {author} {\bibfnamefont {S.}~\bibnamefont
  {Kanno}}\ and\ \bibinfo {author} {\bibfnamefont {J.}~\bibnamefont {Soda}},\
  }\bibfield  {title} {\bibinfo {title} {{Brane world effective action at
  low-energies and AdS / CFT}},\ }\href
  {https://doi.org/10.1103/PhysRevD.66.043526} {\bibfield  {journal} {\bibinfo
  {journal} {Phys. Rev. D}\ }\textbf {\bibinfo {volume} {66}},\ \bibinfo
  {pages} {043526} (\bibinfo {year} {2002})},\ \Eprint
  {https://arxiv.org/abs/hep-th/0205188} {arXiv:hep-th/0205188} \BibitemShut
  {NoStop}%
\bibitem [{\citenamefont {Almheiri}\ \emph {et~al.}(2019)\citenamefont
  {Almheiri}, \citenamefont {Engelhardt}, \citenamefont {Marolf},\ and\
  \citenamefont {Maxfield}}]{Almheiri:2019psf}%
  \BibitemOpen
  \bibfield  {author} {\bibinfo {author} {\bibfnamefont {A.}~\bibnamefont
  {Almheiri}}, \bibinfo {author} {\bibfnamefont {N.}~\bibnamefont
  {Engelhardt}}, \bibinfo {author} {\bibfnamefont {D.}~\bibnamefont {Marolf}},\
  and\ \bibinfo {author} {\bibfnamefont {H.}~\bibnamefont {Maxfield}},\
  }\bibfield  {title} {\bibinfo {title} {{The entropy of bulk quantum fields
  and the entanglement wedge of an evaporating black hole}},\ }\href
  {https://doi.org/10.1007/JHEP12(2019)063} {\bibfield  {journal} {\bibinfo
  {journal} {JHEP}\ }\textbf {\bibinfo {volume} {12}},\ \bibinfo {pages}
  {063}},\ \Eprint {https://arxiv.org/abs/1905.08762} {arXiv:1905.08762
  [hep-th]} \BibitemShut {NoStop}%
\bibitem [{\citenamefont {Penington}(2020)}]{Penington:2019npb}%
  \BibitemOpen
  \bibfield  {author} {\bibinfo {author} {\bibfnamefont {G.}~\bibnamefont
  {Penington}},\ }\bibfield  {title} {\bibinfo {title} {{Entanglement Wedge
  Reconstruction and the Information Paradox}},\ }\href
  {https://doi.org/10.1007/JHEP09(2020)002} {\bibfield  {journal} {\bibinfo
  {journal} {JHEP}\ }\textbf {\bibinfo {volume} {09}},\ \bibinfo {pages}
  {002}},\ \Eprint {https://arxiv.org/abs/1905.08255} {arXiv:1905.08255
  [hep-th]} \BibitemShut {NoStop}%
\bibitem [{\citenamefont {Hung}\ \emph {et~al.}(2011)\citenamefont {Hung},
  \citenamefont {Myers},\ and\ \citenamefont {Smolkin}}]{Hung:2011xb}%
  \BibitemOpen
  \bibfield  {author} {\bibinfo {author} {\bibfnamefont {L.-Y.}\ \bibnamefont
  {Hung}}, \bibinfo {author} {\bibfnamefont {R.~C.}\ \bibnamefont {Myers}},\
  and\ \bibinfo {author} {\bibfnamefont {M.}~\bibnamefont {Smolkin}},\
  }\bibfield  {title} {\bibinfo {title} {{On Holographic Entanglement Entropy
  and Higher Curvature Gravity}},\ }\href
  {https://doi.org/10.1007/JHEP04(2011)025} {\bibfield  {journal} {\bibinfo
  {journal} {JHEP}\ }\textbf {\bibinfo {volume} {04}},\ \bibinfo {pages}
  {025}},\ \Eprint {https://arxiv.org/abs/1101.5813} {arXiv:1101.5813 [hep-th]}
  \BibitemShut {NoStop}%
\bibitem [{\citenamefont {Randall}\ and\ \citenamefont
  {Sundrum}(1999)}]{Randall:1999vf}%
  \BibitemOpen
  \bibfield  {author} {\bibinfo {author} {\bibfnamefont {L.}~\bibnamefont
  {Randall}}\ and\ \bibinfo {author} {\bibfnamefont {R.}~\bibnamefont
  {Sundrum}},\ }\bibfield  {title} {\bibinfo {title} {{An Alternative to
  compactification}},\ }\href {https://doi.org/10.1103/PhysRevLett.83.4690}
  {\bibfield  {journal} {\bibinfo  {journal} {Phys. Rev. Lett.}\ }\textbf
  {\bibinfo {volume} {83}},\ \bibinfo {pages} {4690} (\bibinfo {year}
  {1999})},\ \Eprint {https://arxiv.org/abs/hep-th/9906064}
  {arXiv:hep-th/9906064} \BibitemShut {NoStop}%
\bibitem [{\citenamefont {Gibbons}\ and\ \citenamefont
  {Hawking}(1977)}]{Gibbons:1976ue}%
  \BibitemOpen
  \bibfield  {author} {\bibinfo {author} {\bibfnamefont {G.~W.}\ \bibnamefont
  {Gibbons}}\ and\ \bibinfo {author} {\bibfnamefont {S.~W.}\ \bibnamefont
  {Hawking}},\ }\bibfield  {title} {\bibinfo {title} {{Action Integrals and
  Partition Functions in Quantum Gravity}},\ }\href
  {https://doi.org/10.1103/PhysRevD.15.2752} {\bibfield  {journal} {\bibinfo
  {journal} {Phys. Rev. D}\ }\textbf {\bibinfo {volume} {15}},\ \bibinfo
  {pages} {2752} (\bibinfo {year} {1977})}\BibitemShut {NoStop}%
\bibitem [{\citenamefont {Iwashita}\ \emph {et~al.}(2006)\citenamefont
  {Iwashita}, \citenamefont {Kobayashi}, \citenamefont {Shiromizu},\ and\
  \citenamefont {Yoshino}}]{Iwashita:2006zj}%
  \BibitemOpen
  \bibfield  {author} {\bibinfo {author} {\bibfnamefont {Y.}~\bibnamefont
  {Iwashita}}, \bibinfo {author} {\bibfnamefont {T.}~\bibnamefont {Kobayashi}},
  \bibinfo {author} {\bibfnamefont {T.}~\bibnamefont {Shiromizu}},\ and\
  \bibinfo {author} {\bibfnamefont {H.}~\bibnamefont {Yoshino}},\ }\bibfield
  {title} {\bibinfo {title} {{Holographic entanglement entropy of de Sitter
  braneworld}},\ }\href {https://doi.org/10.1103/PhysRevD.74.064027} {\bibfield
   {journal} {\bibinfo  {journal} {Phys. Rev. D}\ }\textbf {\bibinfo {volume}
  {74}},\ \bibinfo {pages} {064027} (\bibinfo {year} {2006})},\ \Eprint
  {https://arxiv.org/abs/hep-th/0606027} {arXiv:hep-th/0606027} \BibitemShut
  {NoStop}%
\bibitem [{\citenamefont {Kushihara}\ \emph {et~al.}(2021)\citenamefont
  {Kushihara}, \citenamefont {Izumi},\ and\ \citenamefont
  {Shiromizu}}]{Kushihara:2021fbr}%
  \BibitemOpen
  \bibfield  {author} {\bibinfo {author} {\bibfnamefont {K.}~\bibnamefont
  {Kushihara}}, \bibinfo {author} {\bibfnamefont {K.}~\bibnamefont {Izumi}},\
  and\ \bibinfo {author} {\bibfnamefont {T.}~\bibnamefont {Shiromizu}},\
  }\bibfield  {title} {\bibinfo {title} {{Holographic entanglement entropy of a
  de Sitter braneworld with Lovelock terms}},\ }\href
  {https://doi.org/10.1093/ptep/ptab038} {\bibfield  {journal} {\bibinfo
  {journal} {PTEP}\ }\textbf {\bibinfo {volume} {2021}},\ \bibinfo {pages}
  {043E01} (\bibinfo {year} {2021})},\ \Eprint
  {https://arxiv.org/abs/2102.12597} {arXiv:2102.12597 [hep-th]} \BibitemShut
  {NoStop}%
\bibitem [{\citenamefont {Myers}(1987)}]{Myers:1987yn}%
  \BibitemOpen
  \bibfield  {author} {\bibinfo {author} {\bibfnamefont {R.~C.}\ \bibnamefont
  {Myers}},\ }\bibfield  {title} {\bibinfo {title} {{Higher Derivative Gravity,
  Surface Terms and String Theory}},\ }\href
  {https://doi.org/10.1103/PhysRevD.36.392} {\bibfield  {journal} {\bibinfo
  {journal} {Phys. Rev. D}\ }\textbf {\bibinfo {volume} {36}},\ \bibinfo
  {pages} {392} (\bibinfo {year} {1987})}\BibitemShut {NoStop}%
\bibitem [{\citenamefont {Maeda}\ and\ \citenamefont
  {Torii}(2004)}]{Maeda:2003vq}%
  \BibitemOpen
  \bibfield  {author} {\bibinfo {author} {\bibfnamefont {K.-i.}\ \bibnamefont
  {Maeda}}\ and\ \bibinfo {author} {\bibfnamefont {T.}~\bibnamefont {Torii}},\
  }\bibfield  {title} {\bibinfo {title} {{Covariant gravitational equations on
  brane world with Gauss-Bonnet term}},\ }\href
  {https://doi.org/10.1103/PhysRevD.69.024002} {\bibfield  {journal} {\bibinfo
  {journal} {Phys. Rev. D}\ }\textbf {\bibinfo {volume} {69}},\ \bibinfo
  {pages} {024002} (\bibinfo {year} {2004})},\ \Eprint
  {https://arxiv.org/abs/hep-th/0309152} {arXiv:hep-th/0309152} \BibitemShut
  {NoStop}%
\bibitem [{\citenamefont {Cano}(2018)}]{Cano:2018ckq}%
  \BibitemOpen
  \bibfield  {author} {\bibinfo {author} {\bibfnamefont {P.~A.}\ \bibnamefont
  {Cano}},\ }\bibfield  {title} {\bibinfo {title} {{Lovelock action with
  nonsmooth boundaries}},\ }\href {https://doi.org/10.1103/PhysRevD.97.104048}
  {\bibfield  {journal} {\bibinfo  {journal} {Phys. Rev. D}\ }\textbf {\bibinfo
  {volume} {97}},\ \bibinfo {pages} {104048} (\bibinfo {year} {2018})},\
  \Eprint {https://arxiv.org/abs/1803.00172} {arXiv:1803.00172 [gr-qc]}
  \BibitemShut {NoStop}%
\bibitem [{\citenamefont {Padilla}(2003)}]{Padilla:2003qi}%
  \BibitemOpen
  \bibfield  {author} {\bibinfo {author} {\bibfnamefont {A.}~\bibnamefont
  {Padilla}},\ }\bibfield  {title} {\bibinfo {title} {{Surface terms and the
  Gauss-Bonnet Hamiltonian}},\ }\href
  {https://doi.org/10.1088/0264-9381/20/14/315} {\bibfield  {journal} {\bibinfo
   {journal} {Class. Quant. Grav.}\ }\textbf {\bibinfo {volume} {20}},\
  \bibinfo {pages} {3129} (\bibinfo {year} {2003})},\ \Eprint
  {https://arxiv.org/abs/gr-qc/0303082} {arXiv:gr-qc/0303082} \BibitemShut
  {NoStop}%
\bibitem [{\citenamefont {Brown}\ and\ \citenamefont
  {York}(1994)}]{Brown:1993ke}%
  \BibitemOpen
  \bibfield  {author} {\bibinfo {author} {\bibfnamefont {J.~D.}\ \bibnamefont
  {Brown}}\ and\ \bibinfo {author} {\bibfnamefont {J.~W.}\ \bibnamefont {York},
  \bibfnamefont {Jr.}},\ }\bibfield  {title} {\bibinfo {title} {{Microcanonical
  action and the entropy of a rotating black hole}},\ }\href
  {https://doi.org/10.1007/978-94-011-1938-2_3} {\bibfield  {journal} {\bibinfo
   {journal} {Math. Phys. Stud.}\ }\textbf {\bibinfo {volume} {15}},\ \bibinfo
  {pages} {23} (\bibinfo {year} {1994})},\ \Eprint
  {https://arxiv.org/abs/gr-qc/9303012} {arXiv:gr-qc/9303012} \BibitemShut
  {NoStop}%
\bibitem [{\citenamefont {Solodukhin}(1995)}]{Solodukhin:1994yz}%
  \BibitemOpen
  \bibfield  {author} {\bibinfo {author} {\bibfnamefont {S.~N.}\ \bibnamefont
  {Solodukhin}},\ }\bibfield  {title} {\bibinfo {title} {{The Conical
  singularity and quantum corrections to entropy of black hole}},\ }\href
  {https://doi.org/10.1103/PhysRevD.51.609} {\bibfield  {journal} {\bibinfo
  {journal} {Phys. Rev. D}\ }\textbf {\bibinfo {volume} {51}},\ \bibinfo
  {pages} {609} (\bibinfo {year} {1995})},\ \Eprint
  {https://arxiv.org/abs/hep-th/9407001} {arXiv:hep-th/9407001} \BibitemShut
  {NoStop}%
\bibitem [{\citenamefont {Fursaev}\ and\ \citenamefont
  {Solodukhin}(1995)}]{Fursaev:1995ef}%
  \BibitemOpen
  \bibfield  {author} {\bibinfo {author} {\bibfnamefont {D.~V.}\ \bibnamefont
  {Fursaev}}\ and\ \bibinfo {author} {\bibfnamefont {S.~N.}\ \bibnamefont
  {Solodukhin}},\ }\bibfield  {title} {\bibinfo {title} {{On the description of
  the Riemannian geometry in the presence of conical defects}},\ }\href
  {https://doi.org/10.1103/PhysRevD.52.2133} {\bibfield  {journal} {\bibinfo
  {journal} {Phys. Rev. D}\ }\textbf {\bibinfo {volume} {52}},\ \bibinfo
  {pages} {2133} (\bibinfo {year} {1995})},\ \Eprint
  {https://arxiv.org/abs/hep-th/9501127} {arXiv:hep-th/9501127} \BibitemShut
  {NoStop}%
\bibitem [{\citenamefont {Emparan}(2006)}]{Emparan:2006ni}%
  \BibitemOpen
  \bibfield  {author} {\bibinfo {author} {\bibfnamefont {R.}~\bibnamefont
  {Emparan}},\ }\bibfield  {title} {\bibinfo {title} {{Black hole entropy as
  entanglement entropy: A Holographic derivation}},\ }\href
  {https://doi.org/10.1088/1126-6708/2006/06/012} {\bibfield  {journal}
  {\bibinfo  {journal} {JHEP}\ }\textbf {\bibinfo {volume} {06}},\ \bibinfo
  {pages} {012}},\ \Eprint {https://arxiv.org/abs/hep-th/0603081}
  {arXiv:hep-th/0603081} \BibitemShut {NoStop}%
\end{thebibliography}%

\end{document}